\documentclass{article}\usepackage[]{graphicx}\usepackage[]{color}
\makeatletter
\def\maxwidth{ %
  \ifdim\Gin@nat@width>\linewidth
    \linewidth
  \else
    \Gin@nat@width
  \fi
}
\makeatother

\definecolor{fgcolor}{rgb}{0.345, 0.345, 0.345}
\newcommand{\hlnum}[1]{\textcolor[rgb]{0.686,0.059,0.569}{#1}}%
\newcommand{\hlstr}[1]{\textcolor[rgb]{0.192,0.494,0.8}{#1}}%
\newcommand{\hlcom}[1]{\textcolor[rgb]{0.678,0.584,0.686}{\textit{#1}}}%
\newcommand{\hlopt}[1]{\textcolor[rgb]{0,0,0}{#1}}%
\newcommand{\hlstd}[1]{\textcolor[rgb]{0.345,0.345,0.345}{#1}}%
\newcommand{\hlkwb}[1]{\textcolor[rgb]{0.69,0.353,0.396}{#1}}%
\newcommand{\hlkwc}[1]{\textcolor[rgb]{0.333,0.667,0.333}{#1}}%
\newcommand{\hlkwd}[1]{\textcolor[rgb]{0.737,0.353,0.396}{\textbf{#1}}}%

\usepackage{framed}
\makeatletter
\newenvironment{kframe}{%
 \def\at@end@of@kframe{}%
 \ifinner\ifhmode%
  \def\at@end@of@kframe{\end{minipage}}%
  \begin{minipage}{\columnwidth}%
 \fi\fi%
 \def\FrameCommand##1{\hskip\@totalleftmargin \hskip-\fboxsep
 \colorbox{shadecolor}{##1}\hskip-\fboxsep
     \hskip-\linewidth \hskip-\@totalleftmargin \hskip\columnwidth}%
 \MakeFramed {\advance\hsize-\width
   \@totalleftmargin\z@ \linewidth\hsize
   \@setminipage}}%
 {\par\unskip\endMakeFramed%
 \at@end@of@kframe}
\makeatother

\definecolor{shadecolor}{rgb}{.97, .97, .97}
\definecolor{messagecolor}{rgb}{0, 0, 0}
\definecolor{warningcolor}{rgb}{1, 0, 1}
\definecolor{errorcolor}{rgb}{1, 0, 0}
\newenvironment{knitrout}{}{} % an empty environment to be redefined in TeX

\usepackage{alltt}
\usepackage[table]{xcolor}
\definecolor{light-gray}{gray}{0.9}
\usepackage{lmodern}
\usepackage[T1]{fontenc}
\usepackage[utf8]{inputenc}
\usepackage{natbib} % For bibliography
\usepackage{amsmath,amssymb}
\usepackage{colortbl}
\usepackage{array}
\usepackage{graphicx}
\usepackage{color}
\usepackage{datetime}
\usepackage{natbib}
\usepackage{fancyhdr}
\usepackage{multirow}
\usepackage{multicol}
\usepackage{hyperref}
\newcommand*\rot{\rotatebox{90}} % to rotate the labels in table
\usepackage[a4paper, total={6.5in, 10in}]{geometry}

\usepackage{enumerate}
\usepackage{multirow}
\usepackage{xifthen}

\usepackage{palatino} % font

\usepackage{tikz} % pour mind map
\usepackage{adjustbox}
\usepackage[textwidth=4cm]{todonotes}
\usepackage{booktabs} % to be able to use multicolumn in xtable
\usepackage[framemethod=TikZ]{mdframed} % for grey box interesting results 
\mdfdefinestyle{MyFrame}{%
    linecolor=gray!80!white,
    outerlinewidth=1pt,
    roundcorner=10pt,
    innertopmargin=2pt,
    innerbottommargin=5t,
    innerrightmargin=5pt,
    innerleftmargin=5pt,
    backgroundcolor=gray!20!white}

\newcommand*\widebar[1]{%
  \vbox{%
    \hrule height 0.5pt%     % Line above with certain width
    \kern0.5ex%             % Distance between line and content
    \hbox{%
      \kern-0.1em%           % Distance between content and left side of box, negative values for lines shorter than content
      \ifmmode#1\else\ensuremath{#1}\fi%  % The content, typeset in dependence of mode
      \kern-0.1em%      % Distance between content and left side of box, negative values for lines shorter than content
    }% end of hbox
  }% end of vbox
}

\usepackage{setspace}
\usepackage{xifthen}            % for \isempty
\usepackage{array}
\usepackage{amssymb}

\newcommand{\blanco}[1]{  } 
\newcommand{\latin}[1]{\textit{#1}}

\newcommand{\abk}[1]{\mbox{#1}\xdot}
\DeclareRobustCommand\xdot{\futurelet\token\Xdot}
\def\Xdot{%
  \ifx\token\bgroup.%
  \else\ifx\token\egroup.%
  \else\ifx\token\/.%
  \else\ifx\token\ .%
  \else\ifx\token!.%
  \else\ifx\token,.%
  \else\ifx\token:.%
  \else\ifx\token;.%
  \else\ifx\token?.%
  \else\ifx\token/.%
  \else\ifx\token'.%
  \else\ifx\token).%
  \else\ifx\token-.%
  \else\ifx\token+.%
  \else\ifx\token~.%
  \else\ifx\token.%
  \else.\ %
  \fi\fi\fi\fi\fi\fi\fi\fi\fi\fi\fi\fi\fi\fi\fi\fi%
}

\newcommand{\eg}{\abk{\latin{e.\,g}}}

\newlength{\halbebreite}
\DeclareMathOperator{\Nor}{N} % Normal -
\DeclareMathOperator{\se}{se}   % standard error
\newcommand{\partialv}[3][1]{%
\ifthenelse{#1 = 1}{\frac{\partial #2}{\partial #3}}{\frac{\partial^{#1} #2}{\partial #3^{#1}}}
} 

\newcommand{\partials}[3][1]{%
\ifthenelse{#1 = 1}{\frac{d #2}{d #3}}{\frac{d^{#1} #2}{d #3^{#1}}}
} 

\newcommand{\dseps}[2][1]{%
\ifthenelse{#1 = 1}{\frac{d}{d #2}}{\frac{d^{#1}}{d #2^{#1}}}
}

\newcommand{\dsepv}[2][1]{%
\ifthenelse{#1 = 1}{\frac{\partial}{\partial #2}}{\frac{\partial^{#1}}{\partial #2^{#1}}}
}

\newcommand{\ml}[2][1]{% % für Maximum-Likelihood-Schätzer von #1
\ifthenelse{#1 = 1}%
 {\hat{#2}_{\scriptscriptstyle{\mathrm{ML}}}}% 
 {\hat{#2}^{#1}_{\scriptscriptstyle{\mathrm{ML}}}}% z.B. für sigmadach^2
}
\newcommand{\map}[2][0]{% % für MAP-Schätzer von #1
\ifthenelse{#1 = 0}%
 {\hat{#2}_{\scriptscriptstyle{\mathrm{MAP}}}}% 
 {\hat{#2}_{{\scriptscriptstyle{\mathrm{MAP}}_{#1}}}}% z.B. für sigmadach^2
}
\newcommand{\mpm}[2][0]{% % für MPM-Schätzer von #1
\ifthenelse{#1 = 0}%
 {\hat{#2}_{\scriptscriptstyle{\mathrm{MPM}}}}% 
 {\hat{#2}_{{\scriptscriptstyle{\mathrm{MPM}}_{#1}}}}% z.B. für sigmadach^2
}

\newcommand{\simiid}{\mathrel{\overset{\text{iid}}{\thicksim}}} % iid-verteilt 
\newcommand{\given}{\,\vert\,} % für "X gegeben Y" also $X\given Y$ schreiben
\newcommand{\abs}[1]{\left\lvert#1\right\rvert} % Absolutbetrag
\usepackage{calc}
\IfFileExists{upquote.sty}{\usepackage{upquote}}{}
\begin{document}

\title{\bf \huge Power Calculations for Replication Studies}       
\author{Charlotte Micheloud and Leonhard Held\\ 
  Epidemiology, Biostatistics
  and Prevention Institute (EBPI)\\ 
  and Center for Reproducible Science (CRS) \\
  University of Zurich\\ Hirschengraben 84,
  8001 Zurich, Switzerland\\ Email: \texttt{charlotte.micheloud@uzh.ch}}

\maketitle

\begin{center}
\begin{minipage}{12cm}
  \textbf{Abstract}: The reproducibility crisis has led to an increasing number of replication 
studies being conducted. Sample sizes for replication studies are often 
calculated using conditional power based on the effect estimate from the 
original study. 
However, this approach is not well suited as it ignores the uncertainty of the 
original result. Bayesian methods are used  in clinical trials to 
incorporate prior information into power calculations. We propose to adapt 
this methodology to the replication framework and promote the use of predictive 
instead of conditional power in the design of 
replication studies. Moreover, we describe how extensions of the methodology to
sequential clinical trials can be tailored to replication studies. Conditional
and predictive power calculated at an interim analysis are compared and we
argue that predictive power is a useful tool 
to decide whether to stop a replication study prematurely.
A recent project on the replicability of social sciences is used to illustrate
the properties of the different methods.
% 
% The predictive interim power, i.e. the predictive power conditioned on the data
% already collected, is shown  Predictive power generally leads to smaller values than 
% conditional power and does not always increase when increasing the sample size. 
% Adding more subjects to the replication study can in some cases decrease 
% the predictive power. 
\end{minipage}
\end{center}

\section{Introduction}
The replicability of research findings is essential for the credibility of 
science. However, the scientific world is experiencing a crisis 
\citep{Begley2015} as the replicability rate of many fields appears to be 
alarmingly low. As a result, large scale replication projects, where original 
studies  are selected and replicated as closely as possible to the original 
procedures, have been conducted in psychology  \citep{open2015},
social sciences \citep{camerer2018} and economics \citep{camerer2016} among 
others. Replication success is 
usually assessed using significance and $p$-values, compatibility of effect 
estimates, subjective assessments of replication teams and meta-analysis of 
effect estimates \citep[\eg in][]{open2015}. The statistical evaluation of 
replication studies is still generating much discussion and new standards 
are proposed \citep[\eg in][]{Patil2016, Ly2018, held2020}. 

Yet before a replication study is analyzed, it needs to be designed. 
While the conditions of the replication study are ideally identical to the 
original study, the replication sample size stands out as an exception and 
requires further consideration. Using the same sample size as in the original 
study may lead to a severely underpowered replication study, 
even if the effect $\hat\theta_o$ estimated in the original study is the true, 
unknown effect size $\theta$ \citep{good1992}. Standard power calculations 
using the effect estimate from  the 
original study as the basis for the replication study are commonly used. 
% 18.11.2020 - removed to avoid repetition/shorten the paper
% The resulting power 
% is called `conditional' as it is conditioned on an effect estimate which is 
% assumed to be the truth. 
A major criticism of this method is that the uncertainty 
accompanying this original finding is ignored and so the resulting replication 
study is likely to be underpowered \citep{anderson2017}. In this paper, 
we propose alternatives based on predictive power and adapted from Bayesian 
approaches to incorporate prior knowledge to sample size calculation in clinical 
trials \citep{Spieg2004}.
% However, we could not find any replication projects conducted in a proper 
% sequential manner. In order to fill this gap, we propose methods to calculate 
% the interim power, namely the power of a replication study taking into account 
% the data from an interim analysis. Decisions concerning the continuation of the 
% replication study can then be taken based on this interim power. 

% Power calculations for replication study are not trivial and standard methods 
% are often not appropriate.
% This sentence was commented out on March, 30 because seems to be too repetitive 
% and not really useful.
In an era where an increasing number of replication projects are being 
undertaken, optimal allocation of resources appears to be of particular
importance. Adaptive designs are well suited for this purpose and their 
relevance no longer needs to be justified, particularly in clinical trials where
% 18.11.2020 - removed in order to shorten the paper
% Continuing to collect data whenever it is sufficiently clear that the effect 
% we expected is not present is unethical at several levels. In addition to a 
% waste of money, this process also generates a waste of human resources and can 
% have an impact on human or animal lives. In clinical trials for example, 
continuing a study which should be stopped can be a matter of life or death.
% 18.11.2020 - same, trying to shorten it down
% The same applies to preclinical studies, where in addition animals could be 
% reallocated to other experiments and help to tackle the actual threat of 
% underpowered studies in preclinical biomedicine \citep{Neumann2017}. 
Stopping 
for futility refers to the termination of a trial when the data at interim 
indicate that it is unlikely to achieve statistical significance at the end of 
the trial \citep{snapinn2006}. In contrast, stopping for efficacy arises when 
the data at interim are so convincing that there is no need to continue 
collecting more data.
% There exist two major approaches for assessing efficacy and futility. 
% First, the decision to continue or stop the study can be based on efficacy 
% and futility boundaries, which are calculated to control the type I and the 
% type II error rate, respectively \citep{jennison1999}. Several well-known 
% methods of alpha-adjustment are used to make sure that the type I error is 
% not inflated due to multiple testing \citep{pocock1977}. 
% Less attention has been paid in the literature to futility boundaries 
% \citep{schuler2017}.
One approach for assessing efficacy and futility is called stochastic 
curtailment \citep{halperin1982}, where the conditional power of 
the study, given the data so far, is calculated for a range of alternative 
hypotheses. 
% shorten shorten shorten
% The results can then be summarized into plots or tables and help 
% towards the decision of stopping or continuing the data collection.
Instead  of conditional power, predictive power can also be used to judge 
if a trial should be continued \citep{herson1979}. This concept has been 
discussed in depth in \citet{dallow2011} and \citet{rufibach2016}, 
with an emphasis on the choice of the prior in the latter. 

\citet{lakens2014} points out that sequential replication studies could be an
alternative to fixed sample size calculations. This approach has been adopted by
\citet{camerer2018}
in the \emph{Social Science Replication Project} (\emph{SSRP}), a 
large-scale project aiming at evaluating the replicability of social sciences 
experiments published between 2010 and 2015 in \emph{Nature} and \emph{Science}.
A two-stage procedure was used and 21 original studies have been replicated. However, the sequential approach did not include a
power calculation at interim, 
only allowed for a premature stopping for efficacy and did not mention any
adjustment on the threshold
for significance.
% However,  adaptive designs for replication studies is a topic that has not been intensively investigated in the literature.  \citet{camerer2018} conducted the \emph{Social Science Replication Project} (\emph{SSRP}) where they informally performed an interim analysis but without adjusting the threshold for significance or recalculating the power at the interim analysis. Moreover, only stopping for efficacy was considered and no mention of futility stopping has been made.
We try to fill this gap by proposing different methods to calculate the interim 
power, namely the power of a replication study taking into account the data from 
an interim analysis. We argue that \emph{predictive} interim power is a useful 
tool to guide the decision to stop replication studies where the intended effect 
is not present. Our framework only enables power calculation 
at a single interim analysis.

% In this work, the outcomes are assumed to be normally distributed and the effect estimate from the original study to be positive. Moreover, our approach is based on unitless quantities, namely test statistics and relative sample sizes. Such methodology is more intuitive and eases the interpretation.

This paper is structured as follows: power calculations for non-sequential 
(Section~\ref{sec:nonseq}) and sequential (Section~\ref{sec:seq}) replication 
studies are presented, with a focus on comparing conditional and 
predictive methods.
% This paper consists of two main parts: power calculations for non-sequential (Section~\ref{sec:nonseq}) and for sequential (Section~\ref{sec:seq}) replication studies. 
% 28.10.2020 - Really needed?
% We assume normally distributed outcomes.  
Relevant properties of these methods are then illustrated using data from the \emph{SSRP}
in Section~\ref{sec:application}.
%provides a detailed description of this project.
We close with some discussion in Section~\ref{sec:discussion}.

\section{Non-sequential replication studies}\label{sec:nonseq}
Suppose a study has been conducted in order to 
 estimate an unknown effect size $\theta$. We consider the one-sample case 
 throughout this paper but the results can also be generalized to the case 
 of two samples.
 The study produced a positive effect estimate $\hat
\theta_o$. In order to confirm this finding, 
a replication study is planned. Let us assume that the future data of the 
replication study are normally distributed as follows, 
\begin{align*}
Y_{1}, \ldots, Y_{n_r}  \simiid  \Nor\left(\theta, \sigma^2\right) \, ,
\end{align*}
where $\sigma$ is the known standard 
deviation of one observation, assumed to be the same for original and replication 
study.
In the \emph{SSRP}, as well as in most replication projects,
power calculations for the replication studies are based on the original effect 
estimate $\hat\theta_o$.
% The sources for the prior distribution described in the book are for example a single previous study \citep{brown1987}, a meta-analysis of previous results \citep{dersimonian1996} or subjective assessments \citep{Ten1987}. 
In order to incorporate the uncertainty of $\hat\theta_o$ 
% 20.11.2020 shorten shorten shorten
% in the power and sample size calculation of the replication study, 
we use the following 
prior
\begin{equation}
\theta \sim \Nor\left(\hat \theta_o, \sigma^2_o = \sigma^2/n_o\right) \, , \label{eq:prior}
\end{equation}
centered around $\hat\theta_o$
and with variance inversely proportional to the 
original sample size $n_o$ \citep{Spieg2004}.
%
% We denote by $\sigma^2$ the known 
% unit variance from one observation and $\sigma^2_o = \sigma^2/n_o$ is the 
% variance of the estimate $\hat\theta_o$. 
Prior~\eqref{eq:prior} may be too optimistic in practice, 
where original effect estimates tend to be exaggerated \citep{camerer2018}.
This issue and possible solutions are discussed in the 
next section.

In what follows, the different formulas resulting 
from the use of the prior \eqref{eq:prior} are described.
% 20.11.2020 shorten, shorten, shorten
% Proofs and additional 
% information related to this section can be found in 
% Appendix~\ref{sec:nonseq.app}. 
This section is inspired by Section 6.5 in 
\citet{Spieg2004} where Bayesian contributions to selecting the sample size 
of a clinical trial are studied. We adapt this methodology to the replication 
framework and express the power calculation formulas in terms of unitless 
quantities (namely relative sample sizes and test statistics).

% 31.10.2020 : removed as it seems confusing
% This approach has been used in the \emph{SSRP}, 
% where the replication studies were powered to detect 75\% (first data collection)
% and 50\% (second data collection) of the original results. 

\subsection{Methods}\label{sec:methods}
We differentiate between design and  analysis prior,
% shorten shorten shorten
both having an
impact on the power calculation \citep{OHagan2001}, and present the different 
combinations of priors in 
Table \ref{tbl:designanalysis}. 
% 18.11.2020 - shorten shorten shorten 
% The design prior is used 
% before the data are collected in order to quantify prior beliefs about the 
% true effect size \citep{Schonbrodt2018}. It contributes to the study design 
% but is not used in the subsequent statistical analysis. 
% 18.11.2020 - shorten the paper
% The fact that prior 
% information can also be included in the analysis is not well-known. 
% 3.11.2020: Here I could say something like: 
% This is investigated in OHagan where the difference between design 
% and analysis prior is made clear. Incorporating 
% prior (1) as an analysis prior results in a pooling orignal
% and replication studies.

% 3.11.2020: Essai
% A flat analysis prior corresponds to a classical frequentist analysis, 
% while an informative analysis prior indicates a Bayesian analysis 
% \citep{OHagan2001}, where original and replication results are pooled. 

% TABLE %
\newcolumntype{C}{>{\centering\arraybackslash} m{4.5cm} }
\newcolumntype{D}{>{\centering\arraybackslash} m{4.5cm} }
\begin{table}[!h]
\centering
\caption{\small Methods of power calculations resulting from the different combinations of design and analysis priors.}
\begin{tabular}{m{0.7cm} m{2.3cm}|D|C}

\multicolumn{2}{c}{}  & \multicolumn{2}{c}{\textbf{Design}}\\
\multicolumn{2}{c}{}  & \multicolumn{2}{c}{}\\

  &      & Point prior $\theta = \hat\theta_o$ &  Normal prior $\theta \sim \Nor\left(\hat \theta_o, \sigma^2_o\right)$ \\

\cline{2-4}

\multirow{2}{*}{\rot{\textbf{Analysis}}} & Flat prior &   \textcolor{white}
 {blablablablablablablablablablabl} Conditional   \textcolor{white}
 {blablablablablablablablablablabl}  & Predictive\\
\cline{2-4}
& Normal prior  $\theta \sim \Nor\left(\hat \theta_o, \sigma^2_o\right)$ & \textcolor{white}
 {blablablablablablablablablablabl} Conditional Bayesian \textcolor{white}
 {blablablablablablablablablablabl} & Fully Bayesian
\end{tabular}
\label{tbl:designanalysis}
\end{table}

% 18.11.2020 - shorten shorten shorten 
A point prior at $\theta = \hat\theta_o$ in the design
% ignores the uncertainty and 
corresponds to the concept of conditional power  
\citep{SpiegFreed1986}. 
% 9.11.2020 - removed - too repetitive
% In such cases, the probability of a 
% positive result in the replication study is conditioned on the effect estimate 
% from the original study. 
In contrast, the normal design prior \eqref{eq:prior}
% shorten shorten shorten
% incorporates the 
% accompanying uncertainty and 
is related to the concept of predictive power, 
which averages the conditional power over the possible values of the true effect 
according to its design prior distribution. 
% Predictive power is computed by 
% integrating the conditional power with respect to the prior \eqref{eq:prior}. 
% 9.11.2020 - removed 
% \citep{spiegelhalter1986, SpiegFreed1986}. 
Alternative names in the literature are assurance 
\citep{ohagan2005}, probability of study success \citep{wang2013} and Bayesian
predictive power \citep{spiegelhalter1986}. Conditional and predictive power 
are usually accompanied by a flat analysis prior, but can also be calculated 
assuming that original and replication data are pooled (using the
normal analysis prior \eqref{eq:prior}), 
resulting in the conditional Bayesian power and the fully Bayesian power,
respectively.
% Maybe useful? no - shorten shorten shorten 
% A flat analysis prior corresponds to a classical frequentist analysis, 
% while using the analysis prior \eqref{eq:prior} corresponds to
% pooling original and replication data.

In practice, publication bias and the winner's curse often lead to overestimated
original effect estimates \citep{ioannidis2008, button2013, anderson2017}. 
Hence, prior \eqref{eq:prior} might be over-optimistic and 
lead to underpowered replication studies. 
A simple way to correct for this over-optimism 
is to multiply 
the \emph{design} prior mean
$\hat\theta_o$ in \eqref{eq:prior}
by a factor $d$ between $0$ and $1$. The corresponding shrinkage factor 
$s = 1-d$ can be chosen based on previous replication studies in the same field. 
This is the approach considered in the \emph{SSRP} and we expand on this
in Section~\ref{sec:application}. 
More advanced methods 
using empirical Bayes based power estimation \citep{Jiang2016} and data-driven shrinkage
\citep{Pawel2019} are not 
considered here.
% A schematic representation of these four power calculation 
% methods can be found in Figure~\ref{fig:schema1}.
% Appendix~\ref{sec:schema1}.

% 27.10.2020 schematic representation removed 
% according to reviewers comment
% \begin{figure}[!h]
% \centering

% \caption{Schematic representation of power calculations for non-sequential replication studies. The dashed line means that the replication study has not been conducted yet. The estimate $\hat\theta_o$ with black or grey confidence interval indicates whether its uncertainty is taken into account in the power calculation or not, respectively. A classical analysis of the results (flat analysis prior) is indicated by two separates lines for original and replication studies while a continuous line indicates a Bayesian analysis of the results (normal analysis prior).}
% \label{fig:schema1}
% \end{figure}

\subsubsection{Conditional power}

Conditional power is the probability that a replication study will lead to a 
statistically significant conclusion at the two-sided level $\alpha$, given that 
the  alternative hypothesis is true \citep[Section 2.5]{Spieg2004}. 
In the context of a replication study, the alternative hypothesis is 
represented by the effect estimate $\hat\theta_o$ from the original study.

Let $z_{\alpha/2}$ and $\Phi[\cdot]$ respectively denote the $\alpha/2$-quantile and the cumulative distribution function of the standard normal distribution. The conditional power
of a replication study with sample size $n_r$ is
\begin{align}
\text{CP} =  \Phi \left[\frac{\hat\theta_o \sqrt{n_r}}{\sigma} + z_{\alpha/2} 
\right] \label{eq:standardpo} \, ,
\end{align}
see Appendix~\ref{sec:standardpow.app} for a derivation. 
% 22.11.2020 - shorten - sentence not really needed
% The minimum clinically important difference in the context of clinical trials is replaced in \eqref{eq:standardpo} by the effect estimate $\hat\theta_o$ from the original study. 
The required replication sample size $n_r$ can be obtained by rearranging \eqref{eq:standardpo}. 

% 27.10.2020 reviewers comment
A key feature of our framework is 
% the independence of all power/sample size 
% formulas from absolute effect measures. 
that all power/sample size formulas are expressed without 
absolute effect measures.
Simple mathematical rearrangements 
produce an expression which only depends on the original test statistic 
$t_o = \hat\theta_o/\sigma_o = \hat\theta_o\sqrt{n_o}/\sigma$ and the 
variance ratio $c = \sigma^2_o/\sigma^2_r$
% = (\sigma^2/n_o)/(\sigma^2/n_r)$. 
% The variance ratio $c$ 
which
simplifies to the relative sample size $c = n_r/n_o$ 
and represents how much the sample size in the replication study is increased 
as compared to the one in the original study. Formula \eqref{eq:standardpo} 
then becomes
\begin{align}
\text{CP} & = \Phi\left[\sqrt{c}\, t_o + z_{\alpha/2} \right] \, . \label{eq:spow.d}
\end{align}
This formula highlights an intuitive property of the conditional power: 
the larger the evidence in the original study (quantified by $t_o$) or the 
larger the increase in sample size compared to the original study 
(represented by $c$), the larger the conditional power of the replication study.

\subsubsection{Predictive power}

% The previous method calculates the power conditional on the effect from the 
% original study, which is assumed to be the true effect $\theta$. 
% This assumption is questionable, as the original effect estimate $\hat\theta_o$ 
% is an \emph{estimate} of the true effect and hence comes with its uncertainty. 
In order to incorporate the uncertainty of $\hat\theta_o$,
% while still performing a classical analysis, 
the concept of predictive power is discussed \citep{SpiegFreed1986}.
Its formula is:
\begin{align}
\text{PP} =  \Phi\left[\sqrt{\frac{n_o}{n_o+n_r}}\left(\frac{\hat\theta_o 
\sqrt{n_r}}{\sigma}+z_{\alpha/2}\right)\right] \, \label{eq:predpow_or},
\end{align}
see Appendix \ref{sec:predictivepow.app} for a derivation. 
The predictive power \eqref{eq:predpow_or} tends to the conditional power 
\eqref{eq:spow.d} as the original sample size $n_o$ increases. 
Using the unitless 
quantities $t_o$ and $c$, the predictive power can be rewritten as
\begin{align}
\text{PP} & = \Phi\left[\sqrt{\frac{c}{c+1}}\, t_o + \sqrt{\frac{1}{c+1}}\, 
z_{\alpha/2}\right] \, . \label{eq:hpow.d}
\end{align}

\subsubsection{Fully Bayesian and conditional Bayesian power}
So far two power calculation methods where a flat analysis prior is used 
have been considered. 
This approach corresponds to the two-trials rule in drug development, 
which requires ``at least two adequate and well-controlled studies, each 
convincing on its own, to establish effectiveness'' \citep[p. 3]{FDA1998}. 
In practice, this translates to two studies with a significant $p$-value
and an effect in the intended direction.

% replicating the results of a first study in a second, independent
% study and requiring both two-sided $p$-values to be $< \alpha = 0.05$,
% with both effect estimates in the same direction. 

% 22.03.2021 - changes 
% This approach fits the replication 
% framework as a replication study should be an independent repetition of an 
% original study being questioned. 

% An alternative approach for the analysis is to pool original and replication data.
% This is similar to a meta-analysis of original 
% and replication effect estimates, as done in the \emph{SSRP} for example.
% However, in order to ensure the same evidence level as when original and replication studies
% are analysed independently, 
% the corresponding one-sided significance level $\tilde\alpha = \alpha \times \alpha/2 = \alpha^2/2$ should be used 
% \citep{Gibson2020}. \todo{We should add some explanation about why this level is 
% one-sided} This corresponds to an original study at two-sided level $\alpha$ and a
% replication study at one-sided level $\alpha/2$, aiming to replicate the significance
% of the first result, but also its direction.

An alternative approach for the analysis is to pool original and
replication data.
This is similar to a meta-analysis of original
and replication effect estimates, as done in the \emph{SSRP} for
example.
However, in order to ensure the same evidence level as when original
and replication studies
are analyzed independently,
the corresponding two-sided significance level $\tilde\alpha = \alpha^2/2$ should be used
\citep{Fisher1999, Gibson2020}.

% In view of that, a Bayesian analysis where 
% original and replication results are pooled seems less suitable. However, 
% a pooled analysis has interesting properties and is worth a short mention. 

The fully Bayesian power is calculated using the prior \eqref{eq:prior} in 
both the design and the analysis. Using the same prior beliefs in both stages 
is considered as the most natural approach by some authors \citep[\eg in][]{OHagan2001}. The corresponding formula is 

\begin{align}
\text{FBP} = \Phi \left[\sqrt{\frac{c+1}{c}} \, t_o + \sqrt{\frac{1}{c}}\,z_{\tilde\alpha/2}\right] \, . \label{eq:bpow.d}
\end{align}
Note that the fully Bayesian power is also a predictive power as it incorporates the uncertainty of the original effect estimate $\hat\theta_o$.

The last possible combination of design and analysis priors leads
to
% is ignoring the uncertainty of the original result in the design, while still performing a Bayesian analysis. We call the resulting power 
the conditional Bayesian power:
\begin{equation}
\text{CBP} = \Phi\left[\frac{c+1}{\sqrt{c}}\,t_o +\sqrt{\frac{c+1}{c}}\,  z_{\tilde\alpha/2} \right] \, . \label{eq:cbpow.d}
\end{equation}
Derivations of \eqref{eq:bpow.d} and \eqref{eq:cbpow.d} can be found in Appendix \ref{sec:bpow.app} and \ref{sec:cbpow.app}.

% 27.10.2020 - moved to before
% \subsubsection{Inflated original findings}
% The issue of overestimated original findings is widely known and can have several 
% causes \citep{ioannidis2008, button2013, anderson2017}. In the \emph{SSRP}, 
% replication effect estimates were about 50\% of the original effect estimates 
% (see Figure~\ref{fig:SSRP.summary}). A simple way to correct for this inflation 
% in all power calculations is to shrink the original effect estimate $\hat\theta_o$ 
% or equivalently the original test statistic $t_o$ by a factor between $0$ and $1$. 
% This shrinkage approach has been used in the \emph{SSRP}. More advanced methods 
% using a data-driven approach with empirical Bayes  \citep{Pawel2019} are not 
% considered here.

\subsection{Properties}\label{sec:prop}

% 12.10.2020 - This part is removed
% We now discuss properties of the four power formulas described in Section~\ref{sec:methods} using the \emph{SSRP} data. 

% 12.10.2020 This part is now in the Application section
% In Figure~\ref{fig:boxpower2}, the conditional, predictive, conditional Bayesian and fully Bayesian power of the 21 replication studies is shown. The replication sample size considered in the calculations is the one used by the authors in stage 1, ignoring stage 2. To be consistent with the paper, the original effect estimates are shrunken to 75\% of their reported value in the calculations.

% 8.04.2020 I decided to use ni (what they actually used) instead of n75 (what they calculated) so I removed the part below.
%The conditional power in the figure slightly differs from the power calculated by the authors, due to shrinkage on different scales of effect estimates. While the correlation coefficient $r$ is shrunken before Fisher's $z$ transformation in the \emph{SSRP}, we directly shrink the transformed correlation coefficient $z$, where we can assume normality. However, this dissimilarity is not problematic since we are primarily interested in the differences between the four methods rather than reproducing the exact same values as in the paper. 

% New - maybe rephrase and improve

For fixed relative sample size $c$ and two-sided level $\alpha$, all four 
formulas \eqref{eq:spow.d}, \eqref{eq:hpow.d}, \eqref{eq:bpow.d} and 
\eqref{eq:cbpow.d} react to an increase in original test statistic $t_o$ with 
a monotone increase in power.
However, the original result cannot be changed and it is more 
realistic to study the power when varying the relative sample
size $c$ for fixed original test statistic $t_o$ instead.
% 9.11.2020 - try to put somewhere else. 
% The test statistic $t_o$ can easily be transformed into the two-sided $p$-value 
% $p_o = 2\left(1-\Phi(t_o)\right)$.
Consider two original 
studies with $p$-values $0.046$ and $0.005$.
These $p$-values 
correspond to the original studies by 
\citet{Duncan2012} 
% \citet{Derex2013}
and \citet{Shah2012} in the \emph{SSRP} dataset and are used in the following
for illustrative purposes.
% + link the two sentences
% 12.10.2020  removed because text is being rewritten a little.
% The two studies of Table~\ref{tbl:2studies} are then considered in more detail and
% shorten, shorten, shorten
% Their replication power as a function of the relative sample size $c$ is 
% displayed in Figure~\ref{fig:pow2}. 
% We are aware that the range of relative 
% sample sizes $c$
% % shown in Figure~\ref{fig:pow2}
% is extremely wide and relative sample sizes are in practice more likely to be 
% between $c = 0.5$ and $c = 10$. However, some interesting theoretical properties 
% of the different methods can be observed only for very small or very large $c$. 
% In our examples a two-sided 5\% significance level is assumed. 
% 12.10.2020 - This part is now moved to the Application 
% Because our calculations are based on Fisher's $z$-transformed correlation coefficients, the effective sample sizes are used and the relative sample size therefore becomes $c = (n_r -3)/(n_o - 3)$.

% 27.10.2020: removing CBP from figures
% according to reviewers comments.
\begin{figure}[!h]
\centering
\begin{knitrout}
\definecolor{shadecolor}{rgb}{0.969, 0.969, 0.969}\color{fgcolor}
\includegraphics[width=\maxwidth]{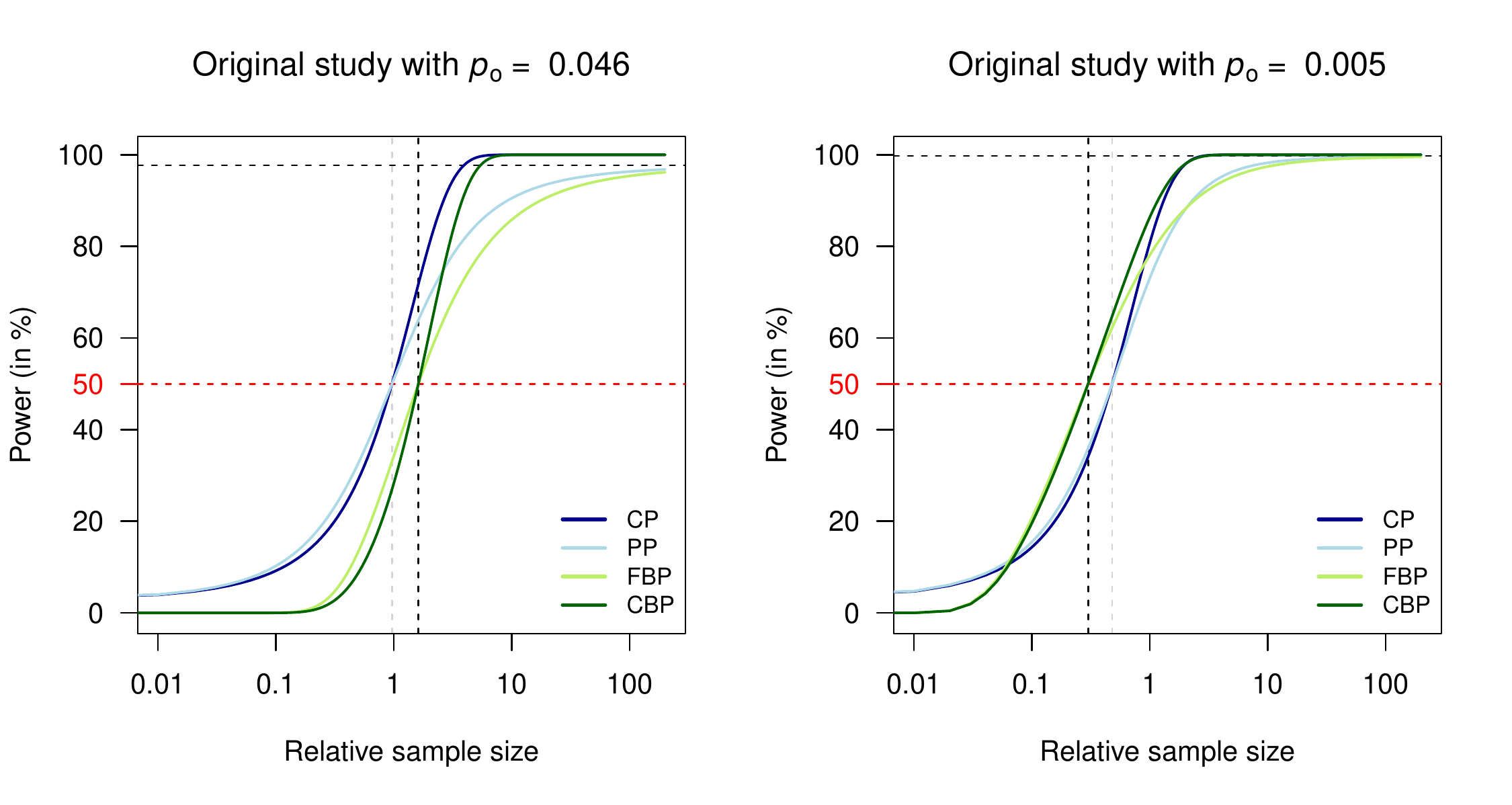} 

\end{knitrout}
 \caption{CP, PP, FBP and CBP as a function of the relative sample size $c$ for 
 two original studies with $p_o =  0.046$ (left) and 
 $p_o =  0.005$ (right) at the two-sided $\alpha = 5$\% level, so
 $\tilde\alpha = 0.00125$.
 The vertical grey line corresponds to the intersection of  
 CP and PP curves as calculated in \eqref{eq:intersec1}, and the vertical 
 black line to the intersection of FBP and CBP as in \eqref{eq:intersec2}. 
 % The horizontal and vertical purple lines indicate the minimum FBP
 % (see \eqref{eq:minimumpower}) and the corresponding relative sample size 
 % $c$ as in~\eqref{eq:cminimumpower}, respectively.
 The horizontal black line 
 indicates the asymptote $1-p_o/2$ of PP and FBP.}
 \label{fig:pow2}
 \end{figure}

\subsubsection{Conditional vs.\ predictive power}\label{sec:condvspred}
% 12.10.2020 - Moved to application

% Figure~\ref{fig:boxpower2} is in agreement with well-known results from the 
% literature \citep{Spieg2004, grouin2007, dallow2011}: predictive methods 
% generally lead to a smaller power than conditional methods. This is the case 
% for all 21 studies of this dataset. 

The power obtained with predictive methods is always closer to 50\% than the 
power obtained with conditional methods \citep{Spieg2004, grouin2007, dallow2011}.
In practice, power is typically larger than 50\% and this implies that 
CP \eqref{eq:spow.d} is larger than  PP \eqref{eq:hpow.d}; and CBP 
\eqref{eq:cbpow.d} is larger than FBP \eqref{eq:bpow.d}. 
% Note that the conditional Bayesian power is not shown 
% in Figure~\ref{fig:boxpower2} as it is very close to 100\% for all the studies. 

% 12.10.2020 - Removed
% However, the situation is reversed when the conditional power is smaller 
% than 50\%. In this case, the predictive power is larger than the conditional 
% power; and the fully Bayesian power larger than the conditional Bayesian power.

% 12.10.2020- Removed (we do not need it anymore)
% Power can also be studied as a function of the relative sample size $c$.

Furthermore, it can be shown that CP and PP are both 
equal to 50\% if the relative sample size is
\begin{equation}
c = z_{\alpha/2}^2/t_o^2 \label{eq:intersec1} \, , 
\end{equation}
the squared $\alpha/2$-quantile of the normal distribution divided by the 
squared test statistic from the original study 
(see Appendix~\ref{sec:condvspred.app} for details). 
Equation~\eqref{eq:intersec1} implies that the larger the evidence in the 
original study (quantified by $t_o$), the smaller the relative sample size $c$ 
where CP and PP curves intersect. 

This can be observed 
in Figure~\ref{fig:pow2}, where the relative sample size at the 
intersection of the CP and PP curves is closer to zero 
in the replication of a convincing original study ($p_o = 
0.005$, $c = 0.48$)
%\citet{Nishi2015} 
 than in the replication of a borderline original study ($p_o = 
0.046$, $c = 0.96$).
 % \citet{Shah2012}. 
 Likewise, FBP and CBP are crossing at a power of 50\% with
 corresponding 
 relative sample size
\begin{equation}
c = z_{\tilde\alpha/2}^2/t_o^2 - 1 \label{eq:intersec2} \, . 
\end{equation} 
% provided the 
% original study is not significant at level $\alpha^2/2$ ($z_{\alpha^2/4}^2 > t_o^2$). Otherwise, 
% the CBP (not shown in Figure~\ref{fig:pow2}) as well as the FBP
% are always larger than 50\%.

\subsubsection{Predictive power cannot always reach 100\%}

Unlike CP \eqref{eq:spow.d} which always reaches 100\% for 
a sufficiently 
large replication sample size, PP \eqref{eq:hpow.d}
has an asymptote at 
$1-p_o/2$. This means that the more convincing the original study, the closer 
to 100\% the PP of an infinitely large replication study is. 
In a sense, the original result penalizes the predictive power.
However, this penalty is not very stringent, as replication of an original 
study with a two-sided $p$-value of 0.05 would still be able to reach a 
PP of 97.5\% for a sufficiently large replication sample size. 
This property also applies to the FBP and can be observed in 
Figure~\ref{fig:pow2} where the horizontal black line indicates
the asymptote $1-p_o/2$. 

\subsubsection{Pooling original and replication studies}
\label{sec:pooling}

% FBP behaves differently depending on the significance status of 
% the original study at level $\alpha^2/2$. For a non-significant original study 
% ($p_o > \alpha^2/2$),
% it converges to 0\% for $c \to 0$ and 
% monotonically increases as the relative sample size $c$ increases. In contrast,
For a borderline significant original study (e.g. $p_o = 0.046$ 
in Figure~\ref{fig:pow2}),  FBP \eqref{eq:bpow.d} and CBP \eqref{eq:cbpow.d} are respectively always smaller
than PP \eqref{eq:hpow.d} and CP \eqref{eq:spow.d}.
In contrast, when the original study is more convincing (e.g. 
$p_o = 0.005$ in Figure~\ref{fig:pow2}), FBP is larger
than PP (respectively CBP larger than CP) for some values of $c$. However, if 
$p_o < \tilde\alpha$, the level required at the end of the replication study
(typically $\tilde\alpha = 0.00125$),
 FBP and CBP converge to 100\% 
for $c \to 0$, decrease down to
\begin{equation}
\Phi\left[\sqrt{t_o^2- z_{\tilde\alpha/2}^2} \,\right] \label{eq:minimumpower}
\end{equation}
for increasing $c$ and then increase to $1 - p_o/2$ (FBP) or 100\% (CBP).
See Appendix~\ref{sec:minBa.app} for a derivation.
A highly convincing original study will thus always have FBP and CBP very 
close to 100\% independently of the sample size. This implies that
a replication may not be required at all, 
a clear disadvantage of pooling original and replication studies instead of 
considering them independently.

\section{Sequential replication studies}\label{sec:seq}
In Section~\ref{sec:nonseq}, power calculations are performed before any data 
have been collected in the replication study. This framework is extended in this
section and allows power (re)calculation at an interim analysis, after some data
have been collected in the replication study already. 
%second round: 16.03.2021
The interim power is defined as the probability of statistical significance at the end 
of the replication study given the data collected so far.
% and a certain assumption about the true effect.
% The interim power is the 
% power of a replication study \emph{taking into account the data from an interim
% analysis}. 
% This implies that independently of the method used, the interim power 
% is conditioned on the data already observed in the replication study. 
% 19.11.2020 shorten shorten shorten  
% If there 
% is none, then the power at interim reduces to the power of non-sequential 
% replication studies as discussed in Section~\ref{sec:nonseq}. 
% The approach of 
% incorporating
The incorporation of prior knowledge into interim power 
has been studied in 
\citet[Section 6.6]{Spieg2004} and we adapt this approach to the case where 
prior information refers to a single original study. Moreover, the power 
calculation formulas are expressed in terms of unitless quantities (relative 
sample sizes and test statistics) in the following.
It is well known from the field of clinical trials that 
the maximum sample size (if the trial has not been stopped at interim) increases with the number of planned interim 
analyses \citep[Section 8.2.1]{mat2006}. In order to maintain a given power, 
even one interim analysis requires a larger maximum sample size than for a trial
with a fixed size and the 
calculation of the replication 
sample size should take this into account.

\subsection{Methods}\label{sec:methods2}
% shorten, shorten, shorten
% Similarly to Section~\ref{sec:methods}, the concepts of design and 
% analysis priors also play a role in this section.

In addition to the point prior $\theta = \hat\theta_o$
and the normal prior \eqref{eq:prior}, the new 
framework enables the specification of a flat design prior. 
% shorten, shorten, shorten
% This refers to a situation where the original study is ignored in the interim power 
% calculation.
Table~\ref{tbl:designanalysis.int} shows the different types of 
interim power calculations that are investigated in this section. 
% and  Appendix~\ref{sec:schema2} provides a schematic representation of each method. 

% TABLE %
\newcolumntype{E}{>{\centering\arraybackslash} m{3cm} }
\newcolumntype{F}{>{\centering\arraybackslash} m{2.6cm} }
\begin{table}[!h]
\centering
\caption{\small Methods of interim power calculations resulting from different 
 design priors using a flat analysis prior.}
\begin{tabular}{E|D|F}
 \multicolumn{3}{c}{\textbf{Design}}\\
 % \multicolumn{3}{c}{}\\
 Point prior $\theta = \hat\theta_o$ &  Normal prior $\theta \sim 
  \Nor\left(\hat \theta_o, \sigma^2_o\right)$ & Flat prior \\
  
\cline{1-3}

  \textcolor{white}
 {blablablablablablablablablablabl} Conditional \textcolor{white}
 {blablablablablablablablablablabl} &
Informed predictive & Predictive 
\end{tabular}
\label{tbl:designanalysis.int}
\end{table}
Calculating the interim power 
to detect
% conditioned on 
the effect estimate from the 
original study ignores the uncertainty of the original result. 
This corresponds to the conditional power in Table~\ref{tbl:designanalysis.int}. 
Uncertainty of the original result can be taken into account  when 
recalculating the power at an interim analysis, turning the conditional power 
into a predictive power. This requires the 
selection of a prior distribution for the true effect, which is updated by the 
data collected so far in the replication study. The prior distributions 
discussed here are the normal prior \eqref{eq:prior} 
(leading to the informed predictive power) and a flat prior 
(leading to the predictive power). The conditional power is then averaged 
with respect to the posterior distribution of the true effect size, given the 
data already observed in the replication study. 
A pooled analysis of original and replication data can also be considered
in this framework
% This framework also enables the specification of an informative prior in the 
% analysis, leading to a pooled analysis .
% However, we know from 
% Section~\ref{sec:nonseq} that this type of analysis is not desired in 
% the present context and 
but is omitted here. 
% Figure~\ref{fig:schema2} shows a schematic representation of the power 
% calculation methods.

% \begin{figure}[!h]
% \centering

% \caption{Schematic representation of power calculations for sequential replication studies. The dashed line represents the fraction of the replication study that has not been conducted yet. The estimate $\hat\theta_o$ with black or grey confidence interval indicates whether its uncertainty is taken into account in the interim power calculation or not, respectively.  A classical analysis of the results (flat analysis prior) is indicated by two separates lines for original and replication studies while a continuous line indicates a Bayesian analysis of the results (normal analysis prior).}
% \label{fig:schema2}
% \end{figure}

Let $\hat\theta_i$ be the effect estimate at interim and 
$\sigma^2_i = \sigma^2/n_i$ the corresponding variance, with $n_i$ the sample 
size at interim. The sample size that is still to be collected in the 
replication study is denoted by $n_j$ and the total replication sample 
size is thus $n_r = n_i + n_j$. The interim power formulas can be shown 
to only depend on the original and interim test statistics $t_o$ and 
$t_i = \hat\theta_i/\sigma_i$, the relative sample size $c = n_r/n_o$ 
and the variance ratio $f = \sigma^2_r/\sigma^2_i = n_i/n_r$, the fraction of
the replication study already completed. 
% Proofs and additional information 
% related to this section can be found in Appendix~\ref{sec:seq.app}. 

\subsubsection{Conditional power at interim}

The conditional power at interim is the interim power to detect the effect 
$\theta = \hat\theta_o$. It can be expressed as

\begin{align}
\text{CPi} & = \Phi\left[\sqrt{c(1-f)}\, t_o  + \sqrt{\frac{f}{1-f}}\, t_i + 
\sqrt{\frac{1}{1-f}}\, z_{\alpha/2} \right] \, , \label{eq:CPis}
\end{align}
see Appendix~\ref{sec:SPi.app} for a derivation. In the particular case where 
no data has been collected yet in the replication study ($f = 0$), the 
CPi \eqref{eq:CPis} reduces to the CP 
\eqref{eq:spow.d}. 
Interim power can also be calculated to detect $\theta = \hat\theta_i$, 
this is however not recommended \citep{bauer2006, Kunzmann2020}.
% Instead of 
% the original effect estimate $\hat\theta_o$, the interim power can be 
% conditioned on 
% the effect estimate $\hat\theta_i$ observed at interim 
% or on a combination of both \citep{bauer2006}. 
% This is however not recommended \citep{dallow2011, Kunzmann2020}.
% Comment: using interim effect estimate is OCP in paper from Kunzmann. 

\subsubsection{Informed predictive power at interim}
The informed predictive power at interim is the predictive interim power using 
the design prior \eqref{eq:prior}. It can be formulated as

\begin{align}
\text{IPPi} & = \Phi \left[\sqrt{\frac{c(1-f)}{(cf+1)(1+c)}}\, t_o + 
\sqrt{\frac{f(1+c)}{(1-f)(cf+1)}}\, t_i + \sqrt{\frac{cf+1}{(1+c)(1-f)}}\, 
z_{\alpha/2} \right] \, , \label{eq:HPPs}
\end{align}
see Appendix~\ref{sec:IPPi.app} for a derivation. 
In the case of $f = 0$ (no data collected in the replication study so far), 
the IPPi \eqref{eq:HPPs} reduces to the PP \eqref{eq:hpow.d}. 
%% Put somewhere else
By considering the original result but also its uncertainty, the predictive 
power at interim is a compromise between considering only the original effect
estimate (CPi) and ignoring the original study completely (PPi).

\subsubsection{Predictive power at interim}

The predictive power at interim is the predictive interim power using a flat design prior. In other words, the results from the original study are ignored. It is expressed as
% \begin{align*}
% \text{PPi} =  \Phi \left(\frac{\sqrt{m+n}}{\sqrt{n}}\frac{\sqrt{m}\hat\theta_i}{\sigma} + \sqrt{\frac{m}{n}}z_{\alpha/2}\right) \, .
% \end{align*}
\begin{align}
\text{PPi} & = \Phi \left[\sqrt{\frac{1}{1-f}}\, t_i + \sqrt{\frac{f}{1-f}}\, z_{\alpha/2} \right] \, , \label{eq:CPPs}
\end{align}
see Appendix~\ref{sec:PPi.app} for a derivation. 
% The predictive power at interim reduces to 0 at the trial start ($f = 0$). As a matter of fact, it is not sensible to use an uninformative design prior when no data has been collected yet. 
Note that PPi \eqref{eq:CPPs} corresponds to FBP
\eqref{eq:bpow.d} provided that the original study in FBP
formula is considered as the interim study (see Appendix \ref{sec:PPi.app}). 
This illustrates the dependence of original and replication studies when a 
normal prior is used in the analysis.
% \subsubsection{Bayesian power at interim}
% This method uses the prior in \eqref{eq:prior} to predict a future Bayesian analysis. The Bayesian Power at interim is
% 
% \begin{align*}
% \text{BPi} =  \Phi \left(\frac{\sqrt{n_0+m+n}}{\sqrt{\left(n_0+m\right)n}}\frac{n_0\hat\theta_o + m \hat\theta_i}{\sigma} + \sqrt{\frac{n_0+m}{n}}z_{\alpha/2} \right) \, .
% \end{align*}
% 
% By setting m to $0$, the Bayesian power at interim is reduced to the Bayesian power at the trial start.
% 
% It can also be expressed as
% \begin{align}
% \text{BPi} & = \Phi\left(\sqrt{1+\frac{c(1-f)}{f(c+1)}}\left(\sqrt{\frac{f}{c(1-f)}}t_o + \sqrt{\frac{f}{1-f}}t_i\right) + \sqrt{\frac{f(c+1)}{c(1-f)}}z_{\alpha/2} \right)\, .
% \end{align}

\subsection{Properties}\label{sec:application2}
Theoretical and specific properties of the conditional, informed predictive
and predictive
power at interim are discussed. 
% \todo[inline]{ADD TEXT HERE}
% 12.10.2020 - part moved to the Application
% The \emph{SSRP} studies are once more used to illustrate the characteristics 
% of the methods introduced above. The studies which have not been stopped at 
% interim have been selected and their interim power has been calculated with 
% different methods (see Table~\ref{tbl:summary_table_interim}). 

% 12.10.2020 - Rewrite this part!!!

% In addition, the informed predictive power at interim and the predictive power at interim are presented in Figure~\ref{fig:6plots} as a function of the fraction ($1-f$) of the replication sample size still to be collected for the two studies from Table~\ref{tbl:2studies} and for two hypothetical interim results. Note that the plots of the predictive power at interim of \citet{Shah2012} and \citet{Nishi2015} are identical as the PPi ignores the original result. A two-sided 5\% significance level is used and effective sample sizes are considered, meaning that $c = (n_r -3)/(n_o -3)$ and $f = (n_i -3)/(n_r-3)$.
% \todo[inline]{Rewrite this part: explain why we focus on when $n_j$ varies
% (because otherwise the interim result also changes, and we are not interested in 
% knowing when to do an interim analysis, but more how does it behave when 
% we increases the sample size still to be collected.)}
% and effective sample sizes are used.
\subsubsection{Conditional vs\, predictive power}\label{sec:condvspredint}

The power at interim, as compared to study start,
involves two additional parameters, namely the test statistic $t_i$ from 
the interim analysis and the fraction $f$ of the replication study already 
conducted.
It is therefore not straightforward to compare the different methods
in terms of which one results in a larger power. Comparison is
facilitated if certain assumptions are made. Consider any combination of a 
significant original result, a non-significant interim result and a replication 
sample size at least twice as large as the original sample size. This translates
to $t_o > z_{1 - \alpha/2}$, 
$t_i < z_{1 - \alpha/2}$ and $c \geq 2$ 
in formulas 
\eqref{eq:CPis}, \eqref{eq:HPPs} and \eqref{eq:CPPs}.
Under these assumptions
and with $f > 0.25$, 
the CPi is always larger than 
the IPPi, which is always larger than the PPi, 
see Appendix \ref{sec:largerthan.app}.
% irrespective of the time of the 
% interim analysis. In contrast, the IPPi
% is larger than the PPi only if $ f > round(cutpoint2$root, 2)$. 
% CPi and PPi can 
% also be compared directly, and the former is larger than the latter
% if $f  > round(cutpoint0$root, 2)$.
% 
% In summary, with the above-mentioned values for $c$, 
% $t_o$ and $t_i$ and 
% with $f > 0.25$, the order is 
% always CPi $\geq$ IPPi $\geq$ PPi.
However, one has to be careful as these conditions are sufficient, 
but not necessary for obtaining this order.

\subsubsection{Weights given to original and interim results}\label{sec:weights}

% Calculating the conditional power at the start of a replication study has been 
% shown to be sub-optimal as the original study often reports an exaggerated 
% effect. 
% 26.10.2020 - not needed?
% In addition to the original result, the conditional power calculated at an 
% interim analysis (CPi) also takes into account the interim result.
% 26.10.2020 - removed. Leo said it has to go before
% The latter, 
% as compared to the former is not subject to publication bias. 
Equations \eqref{eq:CPis}, \eqref{eq:HPPs} and \eqref{eq:CPPs} can be expressed as $\Phi[x]$ where $x$ is a weighted average of $t_o$, 
$t_i$ and $z_{\alpha/2}$ with weights $w_o$, $w_i$ and $w_\alpha$, say.
The weights $w_o$ and $w_i$ depend
on the relative sample size $c$ and the fraction $f$ of the 
replication study already completed.
% (see Appendix~\ref{sec:weights.app})

In the CPi formula \eqref{eq:CPis}, an increase in $c$ leads to a monotone increase in $w_o$ and 
does not affect $w_i$. 
In other words, the weight given to the original result in the CPi becomes 
larger if the relative sample size $c$ increases.
Furthermore, the larger the fraction $f$ of the replication study already 
completed, the less weight is given to the original 
result and conversely, the more weight to the interim result.
% add something here 
% It is also 
% possible to compare the weights given to original and interim
% results as a function of the relative sample size $c$.
% The maximum fraction $f$
% leading to $w_o > w_i$ in
% CPi monotically increases as a function of
% $c$, see Figure~\ref{fig:weights}. 
% In particular, for a replication of the same size as the original  study ($c = 1$), a greater
% weight is given to the original result as compared to the interim result if 
% no more than $f < round(myf2[1], 2)*100$\%. In practice, 
% replication studies are often planned to be larger than their original counterpart
% and larger fractions $f$ of the replication study already completed
% lead the contribution of the original result to be larger than the one of
% the interim result in the conditional power calculation.
% For example, this fraction $f$ can be up to $myf2[2]*100$\% and 
% $round(myf2[3], 2)*100$\% for $c = myc2[2]$ and 
% $c = myc2[3]$, respectively.

In the IPPi formula \eqref{eq:HPPs}, an increase in $f$ leads to a decrease in 
$w_o$ and an increase in $w_i$. Only if the interim analysis takes
place early will  the original result have a greater weight than
the interim result in the 
calculation of the IPPi, 
see Appendix \ref{sec:weights.app}.

% The maximum $f$ so that $w_o > w_i$ is not monotone as a function of $c$. It has
% a maximum value at $f = round(max_myf_ippi, 2)$ for
% $c = max_myc$, and then rapidly decreases and reaches zero in the
% limit. Concretly,

% For example, if $f$ is at most 
% $round(myf2_ippi[1], 2)*100$\%, $round(myf2_ippi[2], 2)*100$\%
% and $round(myf2_ippi[3], 2)*100$\% for $c$ = $myc2[1]$, 
%  $myc2[2]$ and  $myc2[3]$, respectively.
In the PPi formula \eqref{eq:CPPs}, no weight is given to the original result 
and the weight $w_i$ given to interim results increases when $f$ increases.

\subsubsection{A power of 100\% cannot always be reached with the predictive methods}

% 13.10.2020 - new Figure
\begin{figure}[!h]
\centering
\begin{knitrout}
\definecolor{shadecolor}{rgb}{0.969, 0.969, 0.969}\color{fgcolor}
\includegraphics[width=\maxwidth]{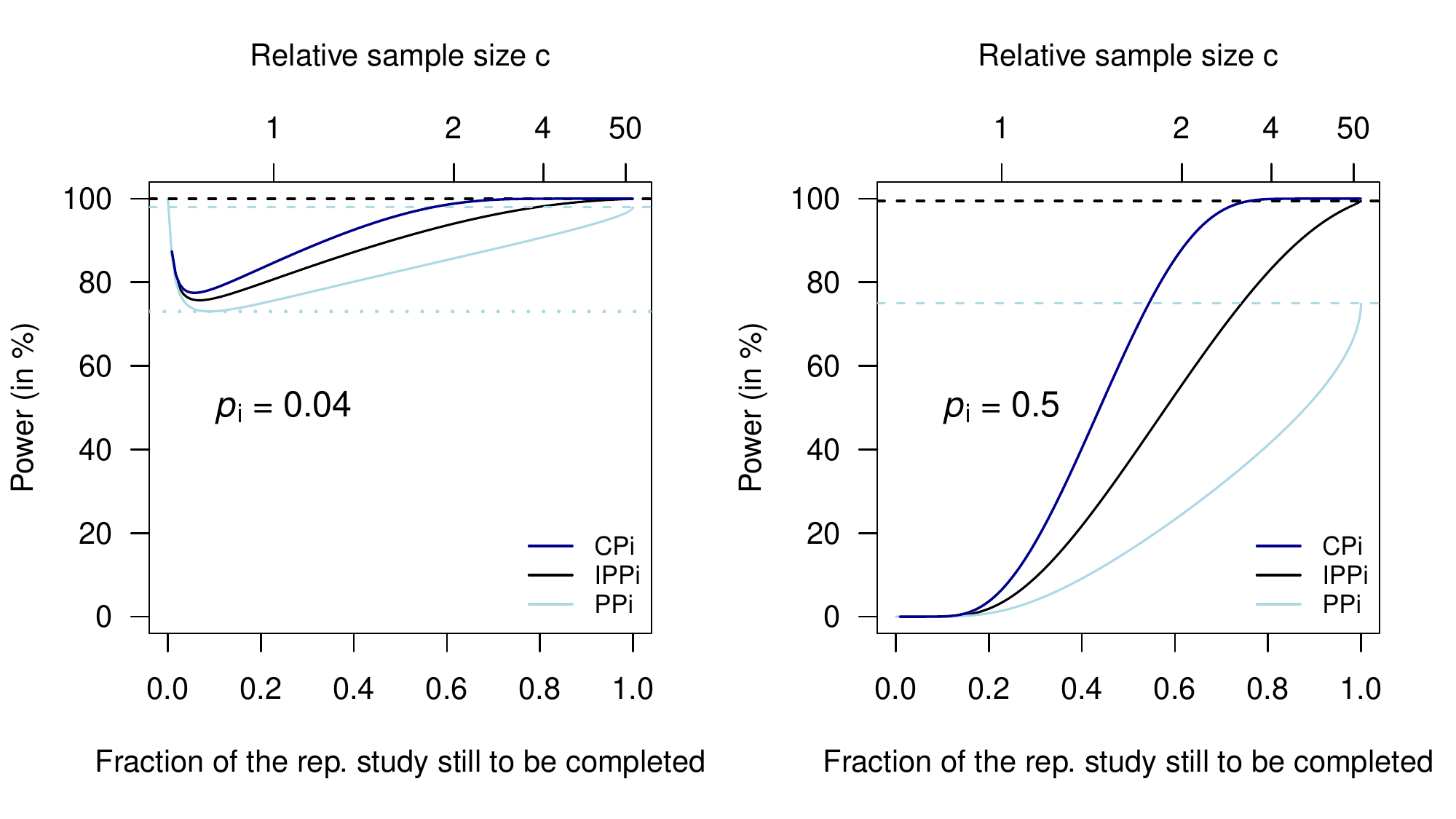} 

\end{knitrout}

\caption{CPi, IPPi and PPi as a function of the sample size $n_j$ still to be collected
in the replication study (or equivalently, as a function of the fraction 
of the replication study still to be completed ($1-f$) and the relative sample
size $c$) for an original study with $p_o = 0.005$
and with two hypothetical interim $p$-values $p_i = 0.04$ (left) 
and $p_i = 0.5$ (right). The two-sided level $\alpha$ is 
$0.05$. Horizontal dashed lines represent the
asymptotes of IPPi and PPi and the horizontal dotted line represents the 
minimum PPi.}
\label{fig:6plots}
\end{figure}

% 13.10.2020 - old plot (from before the revisions)
% \begin{figure}[!h]
% \centering

% \caption{Informed predictive power at interim (IPPi) and predictive power at interim (PPi) as a function of the fraction $1-f$ of the replication study still to be completed for the studies of Table~\ref{tbl:2studies} and with two hypothetical interim $p$-values. Horizontal lines represent asymptotes of IPPi and PPi with $p_i = 0.5$ (black) and $p_i = 0.03$ (red).}
% \label{fig:6plots}
% \end{figure}
Considering that an interim analysis has been conducted, $n_i$ and $t_i$ are 
fixed, and the only parameter that can vary is the sample size $n_j$ 
still to be collected in the replication study. Increasing this sample size 
results in an increase of the relative sample size $c$ and a decrease of the 
fraction $f$ of the replication study already completed.
If $n_j$ is large enough, the CPi \eqref{eq:CPis} reaches 100\%.  In contrast,
% and similarly to the asymptote of predictive power at the start 
% of the replication study
the asymptotes of IPPi \eqref{eq:HPPs} and PPi \eqref{eq:CPPs} are
penalized by the original and/or interim results. The larger the evidence 
in the original study and at interim (represented by $t_o$ and $t_i$, 
respectively), the larger the asymptote of the IPPi
(formula given in Appendix~\ref{sec:asymptotes.app}). The asymptote of 
the PPi, on the other hand, is 
$1-p_i/2$. This last property is explained in
\citet[Section 4]{dallow2011} and the asymptotes can be visualized in 
Figure~\ref{fig:6plots} for an original study with 
$p_o = 0.005$ and two hypothetical interim results: 
$p_i = 0.04$ and $p_i = 0.5$. On the left panel, 
the asymptotes of CPi, IPPi and PPi are all close to 100\% as original and 
interim $p$-values 
are fairly small. A large increase in interim $p$-value hardly has 
an effect on the asymptote of the IPPi (from $99.98$\% 
to $99.5$\%, right panel) but results in a 
dramatic decrease of the asymptote of the PPi and remarkably, 
the maximum PPi achievable for a study with an interim $p$-value of 0.5 is only
$75$\%. 

\subsubsection{Non-monotonicity property of power}\label{sec:non-mono}
If the two-sided interim $p$-value is not significant ($p_i > \alpha$),
the interim power with all three methods behaves in an expected way: 
it increases with increasing 
sample size $n_j$. However, this property breaks when $p_i < \alpha$. 
In this situation, the power assuming no additional subject to be added 
($f = 1$) is 100\%, declines with increasing $n_j$ (decreasing $f$) and then 
increases. For example, the minimum predictive power at interim can be shown 
to be $\Phi\left[\sqrt{t_i^2 - z_{\alpha/2}^2}\,\right]$ which means that the 
PPi of any replication study with a significant interim result will never be 
smaller than 50\% (details in Appendix~\ref{sec:minimumPPi.app}). 
This property can be observed in Figure~\ref{fig:6plots} (left panel)
where the PPi cannot be smaller than $73\%$. 
% The same behavior also applies to the conditional power at interim. 
\citet{dallow2011} explain this characteristic as follows: ``Intuitively, 
if the interim results are very good, any additional subject can be seen as a 
potential threat, able to damage the current results rather than a resource 
providing more power to our analysis.'' 
% No simple analytic expression can be derived for the minimum IPPi. 
% This result implies that if the interim result is significant, 
% this replication study will not to be stopped for futility. 
% Also using this approach to assess an early stopping for efficacy is 
% not optimal as it will always stop if the interim result is significant at 
% level $\alpha$, without taking into account Type I error inflation.

% --- new 30.09: data section comes in the end
\section{Application}\label{sec:application}
Twenty-one significant original findings were replicated in the 
\emph{SSRP} and a two-stage procedure was adopted.
In stage 1, the replication studies had 90\% power to detect 75\% of 
the original effect estimate. Data collection was stopped if a two-sided 
$p$-value $<0.05$ and an effect in the same direction as the original effect 
were found. If not, data collection was continued in stage 2 to have 90\% power 
to detect 50\% of the original effect estimate for the first and second data 
collections pooled. The shrinkage factor $s$ was chosen to be 0.5 as a previous
replication project in the psychological field \citep{open2015} found replication effect estimates
on average half the size of the original effect estimates.
Stages 1 and 2 
can be considered as two steps of a sequential analysis, with an interim 
analysis in between. The analysis after stage 1 will be called 
the \emph{interim} analysis while the \emph{final} analysis will refer to the 
analysis based on the pooled data from stages 1 and 2.

The complete \emph{SSRP} dataset with extended information is available at 
\url{https://osf.io/pfdyw/}. The effects are given as correlation coefficients, 
making them easily interpretable and comparable. Moreover, the application of 
Fisher's $z$ transformation $z(r) = \tanh^{-1}(r)$ to the correlation 
coefficients justifies an asymptotic normal distribution and the standard error 
of the transformed coefficients becomes a function of the effective sample 
size $n-3$ only, $\se(z) = 1/\sqrt{n-3}$. In this dataset, original effects 
are always positive. A ready-to-use dataset \texttt{SSRP} can be found in the 
package \texttt{ReplicationSuccess}, available at 
\url{https://r-forge.r-project.org/projects/replication/}.

\subsection{Descriptive results}
\begin{figure}[!h]
\centering
\begin{knitrout}
\definecolor{shadecolor}{rgb}{0.969, 0.969, 0.969}\color{fgcolor}
\includegraphics[width=\maxwidth]{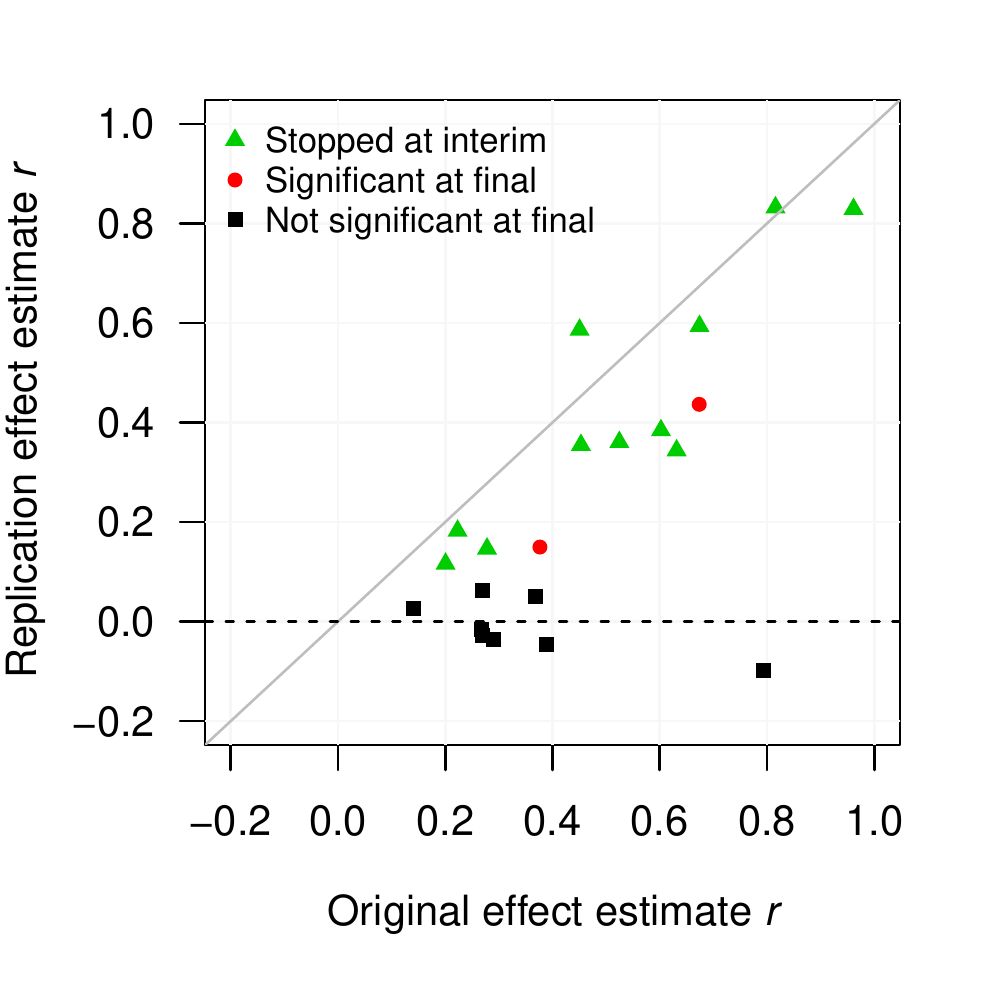} 

\end{knitrout}
\caption{Original effect estimate vs.\ replication effect estimate 
(on the correlation scale). Replications which were not pursued in stage 2 
are included with the results from stage 1. Shape and color of the point 
indicate whether the study was stopped due to a significant result in the 
correct direction at interim (green triangle), was significant
in the correct direction at the final 
analysis (red circle) or was not significant at the final analysis
(black square). The diagonal line indicates replication 
effect estimates equal to original effect estimates. 
% The two studies indicated with an arrow are investigated more 
% extensively in this paper and are described in Table~\ref{tbl:2studies}.
}
\label{fig:SSRP.summary}
\end{figure}

The results are displayed in Figure~\ref{fig:SSRP.summary}.
Twelve studies were significant at interim with an effect in the correct 
direction but by mistake only eleven were stopped. Out of the ten studies 
that were continued, only two showed a significant result in the correct 
direction at the final analysis. The study that was wrongly continued 
turned out to be non-significant at the final analysis. 
The effect of publication bias is clearly seen: 
% in Figure~\ref{fig:SSRP.summary}
original effect estimates are larger than the corresponding replication
effect estimates for 19 out of the 21 studies and are on average twice as large.

% 28.10.2020 - moved to before 
% Aware of the possible inflation of the original effect estimates, the authors 
% of the \emph{SSRP} adopted a two-stage replication sample size calculation 
% procedure. In stage 1, the replication studies had 90\% power to detect 75\% of 
% the original effect estimate. Data collection was stopped if a two-sided 
% $p$-value $<0.05$ and an effect in the same direction as the original effect 
% were found. If not, data collection was continued in stage 2 to have 90\% power 
% to detect 50\% of the original effect estimate for the first and second data 
% collections pooled. This dataset is relevant for our purpose as stages 1 and 2 
% can be considered as two parts of a sequential analysis, with an interim 
% analysis in between. In our framework, the analysis after stage 1 will be called 
% the \emph{interim} analysis while the \emph{final} analysis will refer to the 
% analysis based on the pooled data from stages 1 and 2.

% 19.10.2020: add maybe a comparison with the results.
 \subsection{Power calculations}
 The methods described in Sections~\ref{sec:nonseq} and~\ref{sec:seq} are used
 to calculate the power of the 21 replication studies before
 the onset of the study and at the interim analysis. 
Because our calculations are based on Fisher's $z$-transformed correlation 
coefficients, the effective sample sizes are used. The relative sample size 
is then $c = (n_r -3)/(n_o - 3)$ and the fraction $f$ of the 
replication study already completed $f = (n_i -3)/(n_r-3)$. A two-sided $\alpha =  5$\%
level is used as in the original paper, so $\tilde\alpha = 0.00125$ in 
the calculation of FBP and CBP.
 % \todo[inline]{Here have a short intro and defined c and f. Say: we did two 
 % things (1) reproduce the results and (2) calculate the power at interim}
 %% COMPARISON - power

 \subsubsection{At the replication study start}
\begin{figure}[!h]
\centering
\begin{knitrout}
\definecolor{shadecolor}{rgb}{0.969, 0.969, 0.969}\color{fgcolor}
\includegraphics[width=\maxwidth]{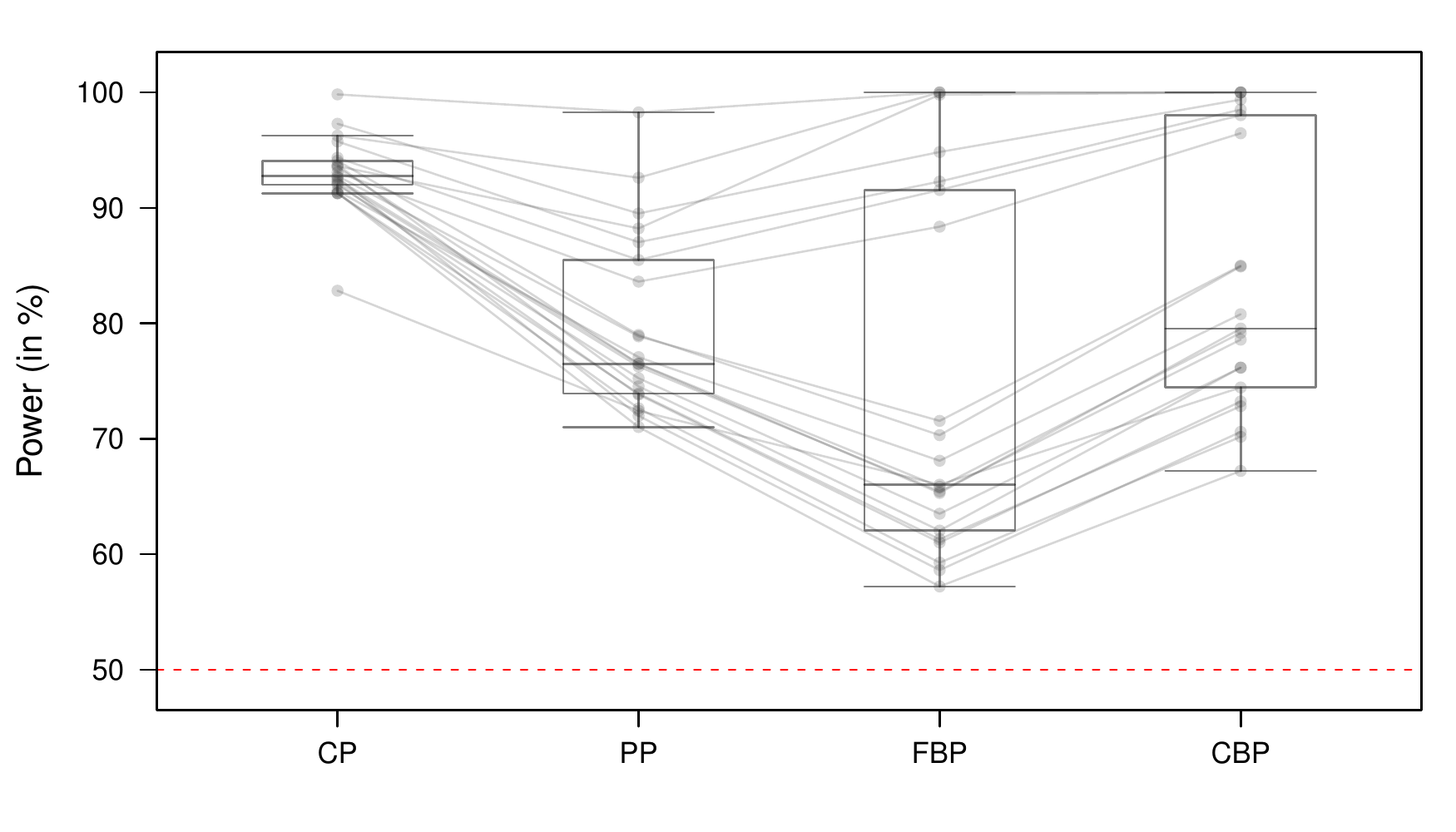} 

\end{knitrout}
\caption{CP, PP, FBP and CBP of the 21 studies of the \emph{SSRP} at level 
$\alpha = 5\%$ (so $\tilde \alpha = 0.00125$ for FBP and CBP)
using a shrinkage
factor $s$ of 0.25 in the calculations.
Each circle represents a study and the lines link the same studies.}
\label{fig:boxpower2}
\end{figure}

We computed the CP, PP, FBP and CBP of the 21 replication studies.
% In Figure~\ref{fig:boxpower2}, the conditional, predictive, 
% conditional Bayesian and fully Bayesian power of the 21 replication 
% studies is shown.
The replication sample size that we considered in the calculations is 
the one used by the authors of the \emph{SSRP} in stage 1, ignoring stage 2. 
To be consistent with the procedures of the \emph{SSRP}, 
a shrinkage factor $s$ of 0.25 was used in the calculations. Results 
can be found in Figure~\ref{fig:boxpower2}, where some 
properties discussed in Section~\ref{sec:prop} are 
illustrated. CP is larger than PP for all studies, and similarly  
CBP is larger than FBP as expected (see Section~\ref{sec:condvspred}). 
Furthermore, it can be oberved that for some 
studies FBP is larger than PP, while it is the opposite for some other studies.
This depends on the $p$-value $p_o$ from the original study and the 
relative sample size $c$ as explained in Section~\ref{sec:pooling}.
The same applies to CP and CBP but cannot be directly observed in 
Figure~\ref{fig:boxpower2}.

% Because our calculations are based on Fisher's $z$-transformed correlation 
% coefficients, the effective sample sizes are used and the relative sample size 
% therefore becomes $c = (n_r -3)/(n_o - 3)$.
% 
% The conditional power is larger than the target 90\% for all except one 
% study, with a median conditional power of 
% $round(median(powers_St), 3)*100$\%. Further explanation about the 
% anomalous study is provided in the 
% Supplementary Information of \citet{camerer2018}. 
% As mentioned in Section~\ref{sec:condvspred}, the 
% predictive power is closer to 50\%, i.e. smaller than the conditional 
% power for the 21 studies. The median predictive power 
% ($round(median(powers_Hy), 3)*100$\%) 
% is considerably smaller than the median conditional power.
% The considerably large fully Bayesian power values are not surprising
% due to all original studies being significant (see Section~\ref{sec:pooling}). 

\subsubsection{At the interim analysis}

% The \emph{SSRP} studies are once more used to illustrate the characteristics 
% of the methods introduced above. 
Replication studies which did not reach significance after the first data 
collection were continued. 
We have selected these studies and calculated their interim power
with the different methods (see Table~\ref{tbl:summary_table_interim}). 
% In addition, the informed predictive power at interim and the predictive power 
% at interim are presented in Figure~\ref{fig:6plots} as a function of the 
% fraction ($1-f$) of the replication sample size still to be collected for the 
% two studies from Table~\ref{tbl:2studies} and for two hypothetical interim results. 
% Note that the plots of the predictive power at interim of \citet{Shah2012} and 
% \citet{Nishi2015} are identical as the PPi ignores the original result. 
% A two-sided 5\% significance level is used and effective sample sizes are 
% considered, meaning that $c = (n_r -3)/(n_o -3)$ and $f = (n_i -3)/(n_r-3)$.
These studies have a  sample size substantially larger in the replication as 
compared to the original study (large $c$). Moreover, 
the interim analysis took place in the second quarter of the 
replication study 
($0.3 \leq f \leq 0.47)$
and by selection, they all have a non-significant interim 
$p$-value (except the study from \citet{Ackerman2010} which was
continued by mistake). Excluding this study, they all fulfill 
the sufficient conditions mentioned in Section~\ref{sec:condvspredint}
and follow the order CPi > IPPi > PPi. This also hold 
for the particular study with a significant interim result 
as the corresponding relative sample size $c$ is large 
($c = 11.62$).

The CPi is remarkably large for all studies, even for the five studies 
where the interim effect estimate is in the opposite direction 
as the original estimate as the weight given to the 
\emph{significant} original
result is consequent due to the large relative sample size $c$
(see Section~\ref{sec:weights}).
% is larger than the weight the interim result in the CPi 
% with this combination of $f$ and $c$, as explained in Section~\ref{sec:weights}.
In contrast, 
% the same values of $f$ and $c$ lead to
more weight is given to the
interim as compared to the original result in the IPPi formula, 
making the corresponding IPPi values more sensible. 
If a futility boundary between 10\% and 30\% had been used 
(as in \citet{demets2006}) four out of the eight studies which 
failed to replicate at the final analysis 
would have been stopped at interim based on the IPPi values.
Surprisingly, the replication study of \citet{Ramirez2011} presents a 
relatively large IPPi ($61.4$\%)
although the interim result goes in the opposite direction 
as the original result. This is due to the very small original $p$-value.
The PPi of the same study 
is considerably smaller ($4.2$\%)
since the original result does not influence the power with this method. 
Furthermore, six out of eight studies which failed to replicate at 
the final analysis would have been stopped at interim if futility stopping 
had been decided based on a PPi of less than 30\%.
Significant interim results 
lead to large PPi values (see Section~\ref{sec:non-mono}),
and that can observed for the study that was incorrectly 
continued.
\begin{table}[ht]
\centering
\caption{\small{CPi, IPPi and PPi of the ten studies that were continued including the original, interim and replication two-sided $p$-values and effect estimates, the relative sample size $c$ and the fraction $f$ of the replication study already completed.}} 
\label{tbl:summary_table_interim}
\scalebox{0.90}{
\begin{tabular}{lccccccccccc}
  \hline
   \multicolumn{1}{c}{} & \multicolumn{2}{c}{Original} & \multicolumn{3}{c}{Interim} & \multicolumn{3}{c}{Interim power} & \multicolumn{3}{c}{Replication} \\  \cmidrule(lr){2-3} \cmidrule(lr){4-6} \cmidrule(lr){7-9} \cmidrule(lr){10-12}
  Study & $p_o$ & $r_o$ & $f$ & $p_i$ & $r_i$ & CPi & IPPi & PPi & $c$ & $p_r$ & $r_r$ \\ 
  \hline
Duncan & 0.005 & 0.67 & 0.37 & 0.29 & 0.18 & 100.0 & 74.6 & 43.4 & 7.42 & 0.00001 & 0.44 \\ 
  Pyc & 0.023 & 0.38 & 0.43 & 0.09 & 0.15 & 100.0 & 85.3 & 71.0 & 9.18 & 0.009 & 0.15 \\ 
  Ackerman & 0.048 & 0.27 & 0.43 & 0.02 & 0.14 & 100.0 & 95.0 & 90.3 & 11.69 & 0.125 & 0.06 \\ 
  Rand & 0.009 & 0.14 & 0.47 & 0.37 & 0.03 & 99.8 & 51.9 & 27.0 & 6.27 & 0.234 & 0.03 \\ 
  Ramirez & 0.000008 & 0.79 & 0.30 & 0.72 & -0.08 & 100.0 & 61.4 & 4.2 & 4.47 & 0.390 & -0.10 \\ 
  Gervais & 0.029 & 0.29 & 0.42 & 0.41 & -0.05 & 97.5 & 1.9 & 0.3 & 9.78 & 0.415 & -0.04 \\ 
  Lee & 0.013 & 0.39 & 0.42 & 0.45 & -0.07 & 97.7 & 3.1 & 0.4 & 7.65 & 0.435 & -0.05 \\ 
  Sparrow & 0.002 & 0.37 & 0.44 & 0.27 & 0.11 & 99.7 & 74.1 & 40.1 & 3.50 & 0.451 & 0.05 \\ 
  Kidd & 0.012 & 0.27 & 0.40 & 0.27 & -0.07 & 98.9 & 1.6 & 0.1 & 8.57 & 0.467 & -0.03 \\ 
  Shah & 0.046 & 0.27 & 0.45 & 0.15 & -0.09 & 87.0 & 0.1 & 0.0 & 11.62 & 0.710 & -0.02 \\ 
   \hline
  \end{tabular}
}
\end{table}

% old - before 27.10.2020

\section{Discussion}\label{sec:discussion}

Conditional power calculations appear to be the norm in most replication 
projects.
% new 29.10.2020 according to reviwer comment
In this paper, we have drawn attention to notable shortcomings 
of this approach and outlined the rationale and properties of 
predictive power. We encourage researchers to abandon conditional methods 
in favor of predictive methods which make a better use of the original 
study and its uncertainty.

% The main goal of this paper is to show that more can and has to be done to ensure
% well-powered replication studies. We discussed the main problem of conditional 
% power and outlined the rationale and properties of predictive power. 
% In contrast to conditional power, increasing sample size does not always 
% increase the predictive power to a desired level. Moreover, because this is 
% an average power, smaller values are usually expected.

Furthermore, as many replications are being conducted and only a fraction 
confirms the original result, we argue for the necessity of sequentially 
analyzing the results. With this in mind, we encourage the initiative from 
\citet{camerer2018} to terminate some replication studies prematurely based 
on an interim analysis. However, their approach only enables efficacy stopping. 
We propose to use interim power to judge if a replication study should be 
stopped for futility. Interim analyses can help to save time and resources 
but also raise new questions with regard to the choice of prior distributions. 
We have shown using studies from the \emph{SSRP} that different design priors 
lead to very different power values and by extension to different decisions. 
Conditioning the power calculations at interim on the original results is even 
more unreasonable than at the study start and leads to very large power values 
given a significant original result, even if interim results suggest evidence 
in the opposite direction.
We recommend the use of IPPi and PPi to make futility decisions.
A 30\% futility boundary is sometimes employed
in clinical trials and has proved to be 
reasonable in the \emph{SSRP}.
Efficacy stopping based on interim power is known to inflate the type-I 
error rate \citep[Chapter 10]{jennison1999}. We only consider futility 
stopping as this issue does not apply here \citep{Lachin2005}.

%% rewrite
% Not only at the study start, but also when power is re-calculated at an interim 
% analysis, the original result should not be taken as it is. 
% %% Here put the emphasis that predictive power should be used
% Predictive power calculations are thus also the preferred alternative at interim.

Some limitations should be noted. 
% First, the paper focuses on power calculations before the onset of the replication 
% study (Section~\ref{sec:nonseq}) to plan the sample size and on power 
% calculations at an interim analysis (Section~\ref{sec:seq}) to guide 
% futility decisions. Advanced methodology for clinical trials
% could be useful to combine the two approaches but this 
% is rarely done in current replication projects such as \emph{SSRP}.
First, the paper discusses power calculations before the onset of the study
and at an interim analysis separately. However, the planned interim analysis
has an impact on power at study start and sample size adjustments are 
necessary \citep[Section 2.1.2]{Wassmer2016}. This is nevertheless rarely 
 done in current replication projects such as \emph{SSRP}.
% Little is said on how to combine the use of the 
% methods from the two sections.
Second, while the ICH E9 `Statistical Principles for Clinical Trials' 
\citep{ICH1999} recommends blinded interim results, our data at interim are 
assumed to be unblinded. This is not a problem for the one-sample case 
but becomes an issue when we want to compare two groups.  
Such a situation would require an Independent Data Monitoring Committee to
prevent the replication study from being biased \citep{kieser2003}. 
% This issue has extensively been covered in the clinical trials literature 
% \citep{kieser2003, bauer2006}. 
% Second, because of the lack of replication projects reporting interim data, 
% the interim power formulas could only be used with few studies. 
% Second, we have shown that interim power is performant for futility decisions 
% but is similar to the approach from \citet{camerer2018} for efficacy decisions. 
Third, 
% only one interim analysis is considered in our approach and 
the assumption of normally distributed observations
is made.

Further research will focus on extending 
this framework to multiple interim analyses in a replication study and to 
sequentially conducted replication studies.
It will also be of interest to apply 
the concept of interim power discussed in Section~\ref{sec:seq} 
to the reverse-Bayes 
assessment of replication success \citep{held2020}.

% As a side note, particular attention should be paid to the nomenclature as it can consistently differ between authors. As an extreme example of incompatibility between nomenclatures, \citet{bauer2006} use the term `predictive power' to describe power conditioned on the interim effect estimate.

 \section*{Software}
 Software for these power calculations can be found in the R-package \\
\texttt{ReplicationSuccess}, available at 
\url{https://r-forge.r-project.org/projects/replication/}. 
An example of the usage of this package is given in Appendix~\ref{sec:package}. 

 \section*{Acknowledgments}
 We thank Samuel Pawel, Malgorzata Roos and Lawrence L. Kupper for helpful comments and suggestions 
 on this manuscript. We also would like to thank the referees whose comments
 helped to improve and clarify the manuscript. 
 \bibliography{biblio}
 
 \appendix

\section{Non-sequential replication study}\label{sec:nonseq.app}

% \subsection{Schematic representation}\label{sec:schema1}
% \begin{figure}[!h]
% \centering

% \caption{Schematic representation of power calculations for non-sequential replication studies.}
% \label{fig:schema1}
% \end{figure}

% new
% \subsection*{Overview}
% \todo[inline]{work in progress}
% \newcolumntype{C}{>{\centering\arraybackslash} m{3.5cm} }
% \newcolumntype{D}{>{\centering\arraybackslash} m{4cm} }
% \begin{table}[!h]
% \centering
% \caption{\small Methods of power calculations resulting from the different 
% combinations of design and analysis priors.}
% \begin{tabular}{m{0.7cm} C|D|D}
% 
% \multicolumn{2}{c}{}  & \multicolumn{2}{c}{\textbf{Predictive distribution}}\\
% \multicolumn{2}{c}{}  & \multicolumn{2}{c}{}\\
% 
%   &      & $\widebar{Y}_{n_r} \sim \Nor\left(\hat\theta_o, \sigma^2/n_r\right)$ 
%   &  $\widebar{Y}_{n_r} \sim \Nor\left(\hat\theta_o, \sigma^2\left(1/n_o + 1/
%   n_r\right)\right)$ \\
% 
% \cline{2-4}
% 
% \multirow{2}{*}{\rot{\textbf{Rej. region}}} & $\widebar{Y}_{n_r} > -\frac{1}
% {\sqrt{n_r}}z_{\alpha/2} \sigma$ & \textcolor{white}
% {blablablablablablablablablablabl} Conditional 
% \textcolor{white}{blablablablablablablablablablabl} & Predictive\\
% \cline{2-4}
% & $\widebar{Y}_{n_r}  >  \frac{-\sqrt{n_o+n_r}z_{\alpha/2} \sigma -
% n_o\hat\theta_o}{n_r}$ &  \textcolor{white}{blablablablablablab} 
% Conditional Bayesian  \textcolor{white}{blablablablablablablablablablabl} & 
% Fully Bayesian
% \end{tabular}
% \label{tbl:overview}
% \end{table}
 \subsection{Conditional power (CP)}\label{sec:standardpow.app}
% 30.10.2020 - moved parts of it in the main text
% 12.10.2020 - shorten down the appendix 
% Suppose researchers conducted a study and declared the results significant at 
% a pre-specified level $\alpha$.  In order to confirm this finding, a replication 
% study is planned. Let us assume that the future data of the replication study are 
% normally distributed, 
% \begin{align*}
% Y_{1}, \ldots, Y_{n_r}  \simiid  \Nor\left(\theta, \sigma^2\right) \, ,
% \end{align*}
% where $\theta$ is the true effect size and $\sigma$ is the known standard 
% deviation of one observation.
Let the sample mean of $Y_{1:n_r}$ be the parameter estimate of the replication 
study with distribution
\begin{align}
\widebar{Y}_{n_r}  \sim  \Nor\left(\theta, \sigma^2/n_r\right) \, . 
\label{eq:likelihoodd.app}
\end{align}
Let us suppose the null hypothesis of the replication study is 
$H_0$: $\theta = 0$ and we want to detect an alternative hypothesis 
$H_1$: $\theta = \hat\theta_o > 0$, where $\hat\theta_o$ is the effect 
estimate of the original study. The corresponding standardized test 
statistic $t_r$ is $\widebar{Y}_{n_r}\sqrt{n_r}/\sigma$ and we declare 
the result statistically significant at the two-sided level  $\alpha$ if
$\abs{t_r} > z_{1-\alpha/2} = - z_{\alpha/2}$. In the following, we 
focus on $t_r > - z_{\alpha/2}$ as $\Pr\left(t_r < z_{\alpha/2}\right)$ is 
relatively 
small for $\hat\theta_o > 0$. 

$H_0$ is rejected if 
\begin{align*}
\widebar{Y}_{n_r} > -\frac{1}{\sqrt{n_r}}z_{\alpha/2} \sigma \, . 
\end{align*}
Following notation from \citet[Section 6.5.2]{Spieg2004}, 
this event is denoted by $S^C_{\alpha/2}$.
% shorten, shorten, shorten - maybe bring back later? 
% and is called the 
% `classical significance'  as a classical 
% (frequentist) analysis will be conducted at the end of the replication study 
% and is opposed to `Bayesian significance' \citep[Section 6.5.3]{Spieg2004} 
% which will become a relevant concept in \ref{sec:bpow.app}. 
% Under $H_1$, $\widebar{Y}_{n_r}$ is normally distributed with mean 
% $\E\left(\widebar{Y}_{n_r}\right) =\hat\theta_o$ and variance 
% $\Var\left(\widebar{Y}_{n_r}\right)=\sigma^2/n_r$.
% In order to calculate the power of the replication study, 
% we compute the probability of `classical significance' given that the effect 
% estimate of the original study is the true effect,
The conditional power can be expressed as
\begin{align}
\Pr\left(S^C_{\alpha/2} \given \theta\right) & =  
\Pr\left( \widebar{Y}_{n_r} > -\frac{1}{\sqrt{n_r}}z_{\alpha/2} \sigma \right) 
\notag \\
% & =  1-\Pr\left(\widebar{Y}_{n_r} \leq -\frac{1}{\sqrt{n_r}}z_{\alpha/2} 
  % \sigma \right) \notag \\
% & =  1- \Pr\left(\frac{\widebar{Y}_{n_r} - \E(\widebar{Y}_{n_r})}
 % {\sqrt{\Var(\widebar{Y}_{n_r})}} \le \frac{-z_{\alpha/2} \sigma/\sqrt{n_r} - 
 % \E(\widebar{Y}_{n_r})}{\sqrt{\Var(\widebar{Y}_{n_r})}} \right) \notag \\
% & =  1 - \Phi\left[\frac{-z_{\alpha/2} \sigma/\sqrt{n_r}  - 
 % \E(\widebar{Y}_{n_r})}{\sqrt{\Var(\widebar{Y}_{n_r})}} \right] \notag\\
% & =  \Phi\left[\frac{z_{\alpha/2} \sigma/\sqrt{n_r}  + \E(\widebar{Y}_{n_r})}
 % {\sqrt{\Var(\widebar{Y}_{n_r})}}\right] \notag \\
% & =  \Phi\left[\frac{z_{\alpha/2} \sigma/\sqrt{n_r}  + \hat\theta_o}
 % {\sqrt{\sigma^2/n_r}}\right] \notag \\
& =  \Phi \left[\frac{\theta \sqrt{n_r}}{\sigma} + z_{\alpha/2} \right] 
 \label{eq:standardpo.app} \, .
\end{align}
% Equation~\eqref{eq:standardpo.app} specifies the conditional power of a 
% replication study with $n_r$ subjects assuming a classical analysis of the 
% results. The necessary sample size to achieve a pre-specified level of power 
% in the replication study can be derived from equation~\eqref{eq:standardpo.app}
% and is given by
% \begin{align*}\label{eq:standardn.app}
% n_r  =  \frac{(z_{1-\beta} - z_{\alpha/2})^2 \sigma^2}{\hat\theta_o^2} \, ,
% \end{align*}
% where $1-\beta$ denotes the power and $z_{1-\beta}$ the ($1-\beta$)-quantile 
% of the standard normal distribution for notational simplicity.
Setting $\theta = \hat\theta_o$ in \eqref{eq:standardpo.app} gives the CP
formula \eqref{eq:standardpo}.
\subsection{Predictive power (PP)}\label{sec:predictivepow.app}
% As a classical analysis is performed at the end of the replication study, 
% we again calculate the probability of `classical significance'. 
% However, 
Predictive power is computed by integrating \eqref{eq:standardpo.app} with 
respect to the prior \eqref{eq:prior}. 
Integration can be demanding and a more direct way is to use the predictive 
distribution of $\widebar{Y}_{n_r}$, 
\begin{equation}
\widebar{Y}_{n_r} \sim \Nor\left(\hat\theta_o, 
 \sigma^2\left(1/n_o + 1/n_r\right)\right) \, , \label{eq:preddistr.app}
\end{equation}
obtained by combining the prior \eqref{eq:prior}
and the likelihood \eqref{eq:likelihoodd.app}. 
The predictive power is
\begin{align*}
\Pr\left(S^C_{\alpha/2}\right) & =  \Pr\left(\widebar{Y}_{n_r} > 
 -\frac{1}{\sqrt{n_r}} z_{\alpha/2} \sigma\right) \\
% \end{align*}
% The design prior is incorporated by using the predictive distribution 
% \eqref{eq:preddistr.app} of $\widebar{Y}_{n_r}$,
% \begin{align*}
% \Pr\left(S^C_{\alpha/2}\right)  & = 1 - \Phi\left[\frac{-z_{\alpha/2} \sigma/
%  \sqrt{n_r} - \E(\widebar{Y}_{n_r})}{\sqrt{\Var(\widebar{Y}_{n_r})}}\right] 
%   \notag \\
% & = \Phi\left[\frac{z_{\alpha/2} \sigma/\sqrt{n_r} + \hat\theta_o}
%  {\sigma \sqrt{1/n_o+1/n_r}} \right] \notag \\
& =  \Phi\left[\sqrt{\frac{n_o}{n_o+n_r}}\left(\frac{\hat\theta_o \sqrt{n_r}}
 {\sigma}+z_{\alpha/2}\right)\right] \, .  
\end{align*}

\subsection{Conditional Bayesian power (CBP)}\label{sec:cbpow.app}

`Bayesian significance'
is denoted as \begin{equation*}
S^B_{\tilde\alpha/2} = \Pr\left(\theta < 0 \given \text{replication data}\right) < 
 {\tilde{\alpha}/2}\, .
\end{equation*}
The event $S^B_{\tilde\alpha/2}$ happens when
\begin{eqnarray*}
\widebar{Y}_{n_r}  >  \frac{-\sqrt{n_o+n_r}z_{\tilde\alpha/2} \sigma - 
 n_o\hat\theta_o}{n_r} \, ,
\end{eqnarray*}
see \citet[Section 6.5.3]{Spieg2004} for details.

The conditional Bayesian power can be expressed as 
\begin{align}
\Pr\left(S^B_{\tilde\alpha/2} \given \theta\right) & 
=\Pr\left(\widebar{Y}_{n_r} > \frac{-\sqrt{n_o+n_r}z_{\tilde\alpha/2} \sigma- 
n_o \hat\theta_o}{n_r}\right) \notag \\
% & =  1-\Phi \left[ \frac{-\sqrt{n_o + n_r} z_{\tilde\alpha/2} \sigma - 
% n_o \hat\theta_o - n_r\E(\widebar{Y}_{n_r})}{n_r \sqrt{\Var(\widebar{Y}_{n_r})}}
% \right] \notag\\
% & = 1- \Phi \left[\frac{-\sqrt{n_o+n_r} z_{\tilde\alpha/2} \sigma - n_o\hat\theta_o - 
% n_r\hat\theta_o}{n_r \sigma/\sqrt{n_r}}\right] \notag \\
& = \Phi \left[
\frac{\hat\theta_o n_o+  \theta n_r}{\sigma\sqrt{n_r}} + 
\sqrt{\frac{n_o + n_r}{n_r}}z_{\tilde\alpha/2} \right] 
\label{eq:condBayesian2.app} \,.
\end{align}
Setting $\theta = \hat\theta_o$ in \eqref{eq:condBayesian2.app} 
gives
\begin{equation}
 \Pr\left(S^B_{\tilde\alpha/2} \given \theta = \hat\theta_o\right) =  \Phi \left[
\frac{\hat\theta_o(n_o + n_r)}{\sigma\sqrt{n_r}} + 
\sqrt{\frac{n_o + n_r}{n_r}}z_{\tilde\alpha/2} \right] \,.
\end{equation}

% Equation \eqref{eq:condBayesian2.app} can be rewritten in terms of $t_o$ and $c$: 
% \begin{equation*}
% \text{CBP} =\Phi\left[\frac{c+1}{\sqrt{c}}\,t_o +\sqrt{\frac{c+1}{c}}\, 
% z_{\tilde\alpha/2} \right] \, . \label{eq:cbpow.d.app}
% \end{equation*}

\subsection{Fully Bayesian power (FBP)}\label{sec:bpow.app}
The fully Bayesian power is computed by integrating \eqref{eq:condBayesian2.app}
with respect to the prior \eqref{eq:prior}. It can be expressed as
\begin{align}
\Pr\left(S^B_{\tilde\alpha/2}\right)  & =    \Pr\left(\widebar{Y}_{n_r} > 
 \frac{-\sqrt{n_o+n_r} z_{\tilde\alpha/2} \sigma- n_o \hat\theta_o}{n_r}\right)
  \notag \\ 
% \end{align*}
% We again incorporate the design prior by using the predictive distribution 
%  \eqref{eq:preddistr.app} of $\widebar{Y}_{n_r}$,
% \begin{align}
%  \Pr\left(S^B_{\tilde\alpha/2}\right) & =  1-\Phi \left[\frac{-\sqrt{n_o + n_r} 
%  z_{\tilde\alpha/2} \sigma - n_o \hat\theta_o - n_r\E(\widebar{Y}_{n_r})}
%  {n_r \sqrt{\Var(\widebar{Y}_{n_r})}}\right] \notag \\
%  & =  1-\Phi\left[\frac{-\sqrt{n_o + n_r} z_{\tilde\alpha/2} \sigma - n_o 
%  \hat\theta_o - n_r\hat\theta_o}{n_r\sigma \sqrt{(n_o + n_r)/ n_r n_o}}\right]
%  \notag \\
& =  \Phi \left[\frac{\hat\theta_o\sqrt{n_o} \sqrt{n_o+n_r}}
{\sigma\sqrt{n_r}} + \sqrt{\frac{n_o}{n_r}}z_{\tilde\alpha/2} \right] .  
\label{eq:bayesianpo.app}
\end{align}
% Equation \eqref{eq:bayesianpo.app} can be rewritten in terms of the original 
% test statistic $t_o = \hat\theta_o/\sigma_o = \hat\theta_o\sqrt{n_o}/\sigma$ 
% and the relative sample size $c = \sigma^2_o/\sigma^2_r = 
% (\sigma^2/n_o)/(\sigma^2/n_r) = n_r/n_o$: 
% \begin{align}
% \text{FBP} = \Phi \left[\sqrt{\frac{c+1}{c}} \, t_o + \sqrt{\frac{1}{c}}\,
% z_{\tilde\alpha/2}\right] \, . \label{eq:bayesianpower.app}
% \end{align}

\subsection{Conditional vs.\ predictive power}\label{sec:condvspred.app}

Conditional (CP) is equal to predictive power (PP) if 
\begin{align*}
&&\Phi \left[ \frac{\hat\theta_o \sqrt{n_r}}{\sigma} + z_{\alpha/2} \right] 
\quad & = \quad \Phi\left[\sqrt{\frac{n_o}{n_o+n_r}} \left(\frac{\hat\theta_o 
 \sqrt{n_r}}{\sigma} + z_{\alpha/2} \right)\right] \notag\\
\Leftrightarrow  &&\frac{\hat\theta_o \sqrt{n_r}}{\sigma} + z_{\alpha/2} 
 \quad \quad & =  \quad \sqrt{\frac{n_o}{n_o+n_r}} \left(\frac{\hat\theta_o 
 \sqrt{n_r}}{\sigma} + z_{\alpha/2} \right) \notag\\
\Leftrightarrow   &&n_r = 0  \quad  &\text{ or} \quad  
 n_r = \frac{\sigma^2 z_{\alpha/2}^2}{\hat\theta_o^2} \\
\Leftrightarrow &&c = 0 \quad  &\text{ or} \quad  c
 = \frac{z_{\alpha/2}^2}{t_o^2} \, . 
\end{align*}
% By plugging  $\sigma^2z_{\alpha/2}^2/\hat\theta_o$ in the conditional 
% (or predictive) power formula, we obtain the replication power corresponding 
% to the intersection of both power curves,
The corresponding CP (or PP) is
\begin{align*}
\Phi \left[ \frac{\hat\theta_o \sqrt{n_r}}{\sigma} + z_{\alpha/2} \right] & =  
\Phi \left[ \frac{\hat\theta_o \sqrt{\sigma^2 z_{\alpha/2}^2/\hat\theta_o^2}}
{\sigma} + z_{\alpha/2} \right] \notag \notag\\
% & =  \Phi \left[ \frac{\hat\theta_o \abs{\sigma z_{\alpha/2}/\hat\theta_o}}
% {\sigma} + z_{\alpha/2} \right]\notag \\
& =  \Phi \left[ \abs{z_{\alpha/2}} + z_{\alpha/2} \right] \notag\\
& =  0.5 \, ,
\end{align*}
because $z_{\alpha/2}$ is always negative and 
$\hat\theta_o$ is assumed to be positive.

Conditional Bayesian (CBP) is equal to fully Bayesian power (FBP) if
\begin{align}
 && \Phi \left[ \sqrt{\frac{n_o + n_r}{n_r}}z_{\tilde\alpha/2} + 
  \frac{\hat\theta_o(n_o+n_r)}{\sigma\sqrt{n_r}}  \right] \quad & = \quad  
  \Phi \left[ \frac{\hat\theta_o\sqrt{n_o} \sqrt{n_o+n_r}}{\sigma\sqrt{n_r}} + 
  \sqrt{\frac{n_o}{n_r}}z_{\tilde\alpha/2} \right] \notag\\
 \Leftrightarrow && \sqrt{\frac{n_o + n_r}{n_r}}z_{\tilde\alpha/2} + 
  \frac{\hat\theta_o(n_o+n_r)}{\sigma\sqrt{n_r}} \quad & = \quad   
  \frac{\hat\theta_o\sqrt{n_o} \sqrt{n_o+n_r}}{\sigma\sqrt{n_r}} + 
  \sqrt{\frac{n_o}{n_r}}z_{\tilde\alpha/2}  \notag\\
 \Leftrightarrow && n_r  \quad & = \quad  \frac{\sigma^2 z_{\tilde\alpha/2}^2 - 
  \hat\theta_o^2 n_o}{\hat\theta_o^2}\notag \\
 \Leftrightarrow &&c \quad & = \quad   \frac{z_{\tilde\alpha/2}^2}{t_o^2} - 
  1\label{eq:bayespseudobayes.app} \, .
\end{align}
By plugging the equation 
\eqref{eq:bayespseudobayes.app} in the FBP (or CBP) formula \eqref{eq:bpow.d},
we once again retrieve a power of $50\%$.

% 20.11.2020 - not really needed
% \subsection{Asymptote of predictive power for $c \to \infty$}
%  \label{sec:asymptpred.app}
% \begin{equation*}
% \lim_{c\to\infty} \text{PP} = \lim_{c\to\infty} \Phi\left[\sqrt{\frac{1}{c+1}}\left(t_o\sqrt{c}+z_{\alpha/2}\right)\right] = \Phi[t_o] \, .
% \end{equation*}

\subsection{Minimum fully Bayesian power}\label{sec:minBa.app}

As the function $\Phi[\cdot]$ is monotonically increasing, it 
is sufficient to consider its argument. The first derivative is
\begin{eqnarray*}
\frac{d}{dn_r}\left(\frac{\hat\theta_o \sqrt{n_o} \sqrt{n_o+n_r}}
 {\sigma\sqrt{n_r}} + \sqrt{\frac{n_o}{n_r}}z_{\tilde\alpha/2} \right) & = &  \notag\\
% \frac{\hat\theta_o\sqrt{n_o}(\frac{1}{2}(n_o+n_r)^{-1/2} \sigma \sqrt{n_r} - 
%  \frac{1}{2} (n_o +n_r)^{1/2})}{\sigma^2 n_r} + \sqrt{n_o}z_{\tilde\alpha/2} n_r^{-3/2}
%   \left(-\frac{1}{2}\right) & = & \notag \\
1/2 \sqrt{n_o}n_r^{-3/2}\left(\frac{\hat\theta_o n_r(n_o+n_r)^{-\frac{1}{2}}}
{\sigma} - \frac{\hat\theta_o (n_o +n_r)^{1/2}}{\sigma} - z_{\tilde\alpha/2} \right) 
\, . \label{eq:bayesderiv.app}
\end{eqnarray*}
By setting it to $0$ and solving for $n_r$, we obtain the replication sample 
size $n_r$ needed to reach the minimum power, which turns out to be
\begin{eqnarray*}
&&  \frac{1}{2} \sqrt{n_o}n_r^{-\frac{3}{2}}\left(\frac{\hat\theta_o n_r
 (n_o+n_r)^{-\frac{1}{2}}}{\sigma} - \frac{\hat\theta_o (n_o +n_r)^
 {\frac{1}{2}}}{\sigma} - z_{\tilde\alpha/2} \right) = 0  \\
 \Leftrightarrow  && n_r  =   n_o \left(\frac{\hat\theta_o^2 n_o}
 {\sigma^2 z_{\tilde\alpha/2}^2} - 1\right) \notag \\ 
  \Leftrightarrow && c = \frac{t_o^2}{z_{\tilde\alpha/2}^2} -1 \, . 
 \label{eq:nmaxpow.app}
\end{eqnarray*}
% By plugging \eqref{eq:nmaxpow.app} in the alternative fully Bayesian power 
% formula given in \eqref{eq:bayesianpower.app}, we find 
The corresponding minimum fully Bayesian power is
\begin{equation*}
\Pr(S^B_{\tilde\alpha/2})  =  \Phi\left[\sqrt{t_o^2 - z_{\tilde\alpha/2}^2}\, \right]\, . 
\end{equation*}

\section{Sequential replication studies}\label{sec:seq.app}

% \subsection{Schematic representation}\label{sec:schema2}

% \begin{figure}[!h]
% \centering

% \caption{Schematic representation of power calculations for sequential replication studies.}
% \label{fig:schema2}
% \end{figure}

% \subsection*{Overview}
% \todo[inline]{work in progress}
% % TABLE %
% \newcolumntype{E}{>{\centering\arraybackslash} m{4.2cm} }
% \newcolumntype{F}{>{\centering\arraybackslash} m{2.6cm} }
% \begin{table}[!h]
% \centering
% \caption{\small Methods of interim power calculations resulting from different 
% combinations of design and analysis priors.}
% \begin{tabular}{m{0.7cm} E|F|E|F}
% 
% \multicolumn{2}{c}{}  & \multicolumn{3}{c}{\textbf{Predictive distribution}}\\
% \multicolumn{2}{c}{}  & \multicolumn{3}{c}{}\\
% 
%   &      & $\widebar{Y}_{n_j} \sim \Nor\left(\hat\theta_o, \sigma^2/n_r\right)$ 
%   &  $\widebar{Y}_{n_j} \given \hat\theta_i \sim \Nor\left(\frac{n_o\hat\theta_o 
%   + n_i\hat\theta_i}{n_o + n_i}, \sigma^2\left(\frac{1}{n_o + n_i} + \frac{1}
%   {n_j}\right)\right)$ & Flat \\
% 
% \cline{2-5}
% 
% \multirow{2}{*}{\rot{\textbf{Rej. region}}} & $\widebar{Y}_{n_j} > 
% \frac{-\sqrt{n_i+n_j}z_{\alpha/2}\sigma - n_i\hat\theta_i}{n_j}$  & 
% \textcolor{white}{blablablablablablablablablablabl} Conditional 
% \textcolor{white}{blablablablablablablablablablabl} & Informed predictive
% & Predictive \\
% \cline{2-5}
% & $\widebar{Y}_{n_j}  >  \frac{-\sqrt{n_o+n_i+n_j}z_{\alpha/2} \sigma - 
% (n_o\hat\theta_o + n_i\hat\theta_i)}{n_j}$ & -  & Fully Bayesian & - \\
% \end{tabular}
% \label{tbl:overview2}
% \end{table}

\subsection{Conditional power at interim (CPi)}\label{sec:SPi.app}
%% 20.11.2020 - shorten, this is repetition
% The following notation is used: $n_o$ and $\hat\theta_o$ define the sample size 
% and effect estimate in the original study, $n_i$ and $\hat\theta_i$ the sample 
% size and the effect estimate at the interim analysis, $n_j$ the number of 
% additional subjects to be collected in the replication study and 
$\widebar{Y}_{n_j}$ is the parameter estimate corresponding to the fraction 
of the replication study still to be completed. 
% The common standard deviation of one observation is denoted 
% by $\sigma$ and is assumed to be known and the same in the original 
% and the replication study. 
Since 
\begin{equation*}
\frac{n_i\hat\theta_i + n_j\widebar{Y}_{n_j}}{n_i+n_j} \sim \Nor \left(\theta, \frac{\sigma^2}{n_i+n_j}\right) \, ,
\end{equation*}
a significant result is found at then end of the replication study if
\begin{equation*}
\widebar{Y}_{n_j} > \frac{-\sqrt{n_i+n_j}z_{\alpha/2}\sigma - 
n_i\hat\theta_i}{n_j} \, , 
\end{equation*}
denoted by $S^C_{\alpha/2}$. Additional details are provided in \citet[Section 6.6.3]{Spieg2004}.
The conditional power at interim is 
\begin{align}
\Pr\left(S^C_{\alpha/2} \given \hat\theta_i, \theta\right) & = 
\Pr\left(\widebar{Y}_{n_j} > \frac{-\sqrt{n_i+n_j}z_{\alpha/2}\sigma - 
n_i\hat\theta_i}{n_j}\right) \notag \\
& = \Phi\left[\frac{\sqrt{n_j}\theta}{\sigma} + \frac{n_i 
\hat\theta_i}{\sigma \sqrt{n_j}} + \sqrt{\frac{n_i+n_j}{n_j}}
z_{\alpha/2}\right] \label{eq:SPi.app}
\end{align}
Setting $\theta = \hat\theta_o$ in \eqref{eq:SPi.app} gives
\begin{align}
\Pr\left(S^C_{\alpha/2} \given \hat\theta_i, \theta = \hat\theta_o\right) = 
\Phi\left[\frac{\sqrt{n_j}\hat\theta_o}{\sigma} + \frac{n_i \hat\theta_i}{\sigma \sqrt{n_j}} + \sqrt{\frac{n_i+n_j}{n_j}}z_{\alpha/2}\right].
\end{align}
% shorten, shorten, shorten
% Let the original and interim test statistics be $t_o = \hat\theta_o/\sigma_o =
% \hat\theta_o \sqrt{n_o}/\sigma$ and $t_i = \hat\theta_i/\sigma_i = \hat\theta_i
% \sqrt{n_i}/\sigma$, the relative sample size  $c = n_r/n_o = (n_i +n_j)/n_o$ 
% and the fraction of the replication study completed so far be $f = n_i/n_r = 
% n_i/(n_i+n_j)$. Then, Equation~\eqref{eq:SPi.app} can be rewritten in terms of
% $t_o$, $t_i$, $c$ and $f$: 
% \begin{align}
% \text{CPi} & = \Phi\left[\sqrt{c(1-f)}\, t_o  + \sqrt{\frac{f}{1-f}}\, t_i + \sqrt{\frac{1}{1-f}}\, z_{\alpha/2} \right] \, . \label{eq:SPi2.app}
% \end{align}

\subsection{Informed predictive power at interim (IPPi)}\label{sec:IPPi.app}
Informed predictive power at interim is computed by integrating \eqref{eq:SPi.app}
with respect to the posterior distribution obtained by combining
the prior \eqref{eq:prior} and the data collected so far in the replication study. 
The IPPi can be expressed as
% Using the predictive distribution of $\widebar{Y}_{n_j}$ in \eqref{eq:preddistr.app}, it turns out that 
\begin{align}
\Pr\left(S^C_{\alpha/2} \given \hat\theta_i, \mbox{prior } \eqref{eq:prior}\right) =  \Phi \left[\sqrt{\frac{n_o n_j}{(n_o+n_i)(n_o+n_i+n_j)}}\frac{\sqrt{n_o}\hat\theta_o}{\sigma} + \sqrt{\frac{n_i\left(n_o+n_i+n_j\right)}{n_j\left(n_o+n_i\right)}} \frac{\sqrt{n_i}\hat\theta_i}{\sigma} \right.\notag \\
\left.  + \sqrt{\frac{\left(n_i+n_j\right)\left(n_o+n_i\right)}{n_j\left(n_o+n_i+n_j\right)}}z_{\alpha/2}\right] \,  . \label{eq:ippi.app}
\end{align}
% which can be rewritten as 
% \begin{align}
% \text{IPPi} & =\Phi \left[\sqrt{\frac{(1-f)c}{(cf+1)(1+c)}}\, t_o + 
% \sqrt{\frac{f(1+c)}{(1-f)(cf+1)}}\, t_i + \sqrt{\frac{cf+1}{(1+c)(1-f)}}\, 
% z_{\alpha/2} \right] \, . \label{eq:HPPs.app}
% \end{align}

\subsection{Predictive power at interim (PPi)}\label{sec:PPi.app}
The PPi formula can be  found by setting $n_o = 0$ in \eqref{eq:ippi.app}, 
which results in
\begin{align}
\Pr\left(S^C_{\alpha/2} \given \hat\theta_i \right) =  \Phi \left[\sqrt{\frac{n_i+n_j}{n_j}}\frac{\sqrt{n_i}\hat\theta_i}{\sigma} + \sqrt{\frac{n_i}{n_j}}z_{\alpha/2}\right] \, . \label{eq:PPi.app}
\end{align}
% Equivalently, 
% \begin{align*}
% \text{PPi} & = \Phi \left[\sqrt{\frac{1}{1-f}}\, t_i + \sqrt{\frac{f}{1-f}}\, z_{\alpha/2} \right] \, . 
% \end{align*}
Equation \eqref{eq:PPi.app} is identical to \eqref{eq:bayesianpo.app} given that $n_i = n_o$, $n_j = n_r$ and $\hat\theta_i = \hat\theta_o$.

\subsection{The ordering of different types of interim power}\label{sec:largerthan.app}

The CPi \eqref{eq:CPis} is greater than (or equal to) the IPPi \eqref{eq:HPPs}
if 
\begin{align}
\sqrt{c(1-f)}\left(1 - \frac{1}{\sqrt{(cf +1)(1 + c)}}\right) t_o \notag
+ \sqrt{\frac{f}{1-f}}\left(1 - \sqrt{\frac{1+c}{cf + 1}}\right) t_i \notag \\[6pt]
+ \frac{1}{\sqrt{1-f}}\left( 1 - \sqrt{\frac{cf + 1}{1+c}}\right) z_{\alpha/2} \geq 0 \, . \label{eq:cpiippi}
\end{align}

If $t_o > z_{1 - \alpha/2}$, 
$t_i < z_{1-\alpha/2}$ and $c > 2$, 
equation \eqref{eq:cpiippi} holds for every $0 < f < 1$.
\\~\\
The IPPi \eqref{eq:HPPs} is greater than (or equal to) the PPi \eqref{eq:CPPs} if 

\begin{eqnarray}
\sqrt{\frac{(1-f)c}{(cf+1)(1 +c)}}t_o 
+ \frac{1}{\sqrt{1-f}}\left(\sqrt{\frac{f(1+c)}{cf + 1}} - 1 \right)t_i \notag \\[6pt]
+ \frac{1}{\sqrt{1 - f}}\left(\sqrt{\frac{cf + 1}{1+c}} - 
\sqrt{f}\right)z_{\alpha/2} \geq 0 \, .\label{eq:ippippi} 
\end{eqnarray}

If $t_o > z_{1 - \alpha/2}$, 
$t_i < z_{1-\alpha/2}$ and $c > 2$, 
equation \eqref{eq:ippippi} holds for $0.25 < f < 1$.

\subsection{Weights given to original and interim results}\label{sec:weights.app}
\begin{figure}[!h]
 \centering
\begin{knitrout}
\definecolor{shadecolor}{rgb}{0.969, 0.969, 0.969}\color{fgcolor}
\includegraphics[width=\maxwidth]{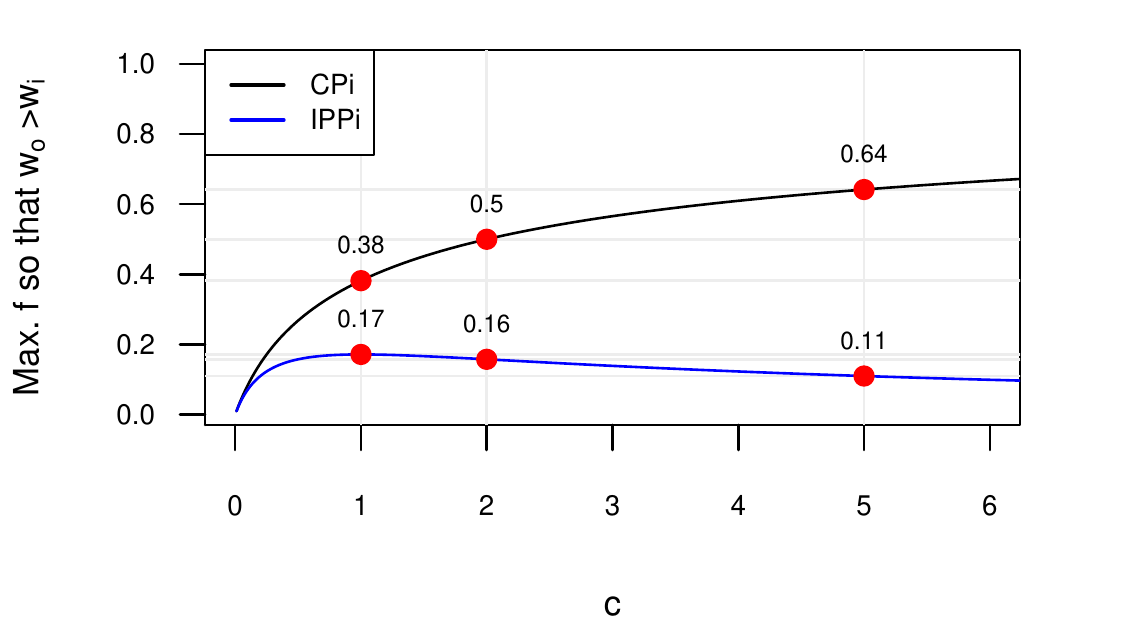} 

\end{knitrout}
\caption{Maximum fraction $f$ of the replication study already completed so 
that $w_o > w_i$ in formulas \eqref{eq:CPis} (CPi) and \eqref{eq:HPPs} 
(IPPi).}
\label{fig:weights}
\end{figure}

In the CPi formula \eqref{eq:CPis}, $w_o > w_i$ if 
\begin{eqnarray*}
 && \sqrt{c(1-f)} > \sqrt{f/(1-f)} \\[6pt]
\Leftrightarrow && f < 1 - \frac{\sqrt{4c + 1} - 1}{2c} \, .
\end{eqnarray*}
In the PPi formula \eqref{eq:HPPs}, $w_o > w_i$ if

\begin{eqnarray*}
&& \sqrt{\frac{c(1-f)}{(cf+1)(1+c)}} > \sqrt{\frac{f(1+c)}{(1-f)(cf+1)}} \\[6pt]
\Leftrightarrow && f <\frac{c^2 - \sqrt{c^2 + 6c + 1}c - \sqrt{c^2 + 6c + 1} + 4c + 1}{2c} \,.
\end{eqnarray*}
Figure~\ref{fig:weights} shows the conditions on $f$ so that $w_o > w_i$ as 
a function of $c$.
% \subsubsection{CPi}
% In the CPi formula in \eqref{eq:SPi2.app}, more weight is given to the original result if 
% \begin{align*}
% && \sqrt{c(1-f)} & > \sqrt{\frac{f}{1-f}}\\
% \Leftrightarrow && c(1-f) & > \frac{f}{1-f} \\
% \Leftrightarrow && c & > \frac{f}{(1-f)^2}
% \end{align*}
% 
% \subsubsection{IPPi}
% In the IPPi formula in \eqref{eq:HPPs.app}, more weight is given to the original result if 
% 
% \begin{align*}
% && \sqrt{\frac{(1-f)c}{(cf+1)(1+c)}} & > \sqrt{\frac{f(1+c)}{(1-f)(cf+1)}} \\
% \Leftrightarrow && \frac{c}{(c+1)^2} & > \frac{f}{(1-f)^2}
% \end{align*}

\subsection{Asymptote of IPPi}\label{sec:asymptotes.app}
For $n_j \to \infty$, the limiting value of the IPPi is
\begin{align*}
% \lim_{n_j \to \infty} \text{CPi} & = \infty \\
\lim_{n_j \to \infty} \text{IPPi} & = \Phi\left[\sqrt{\frac{1}{n_i/n_o + 1}}\, 
t_o + \sqrt{\frac{n_i/n_o}{n_i/n_o + 1}}\,  t_i \,\right] \, .
% \lim_{n_j \to \infty} \text{PPi} & = \Phi\left[t_i\right]
\end{align*}

\subsection{Minimum PPi}\label{sec:minimumPPi.app}
As the PPi formula in \eqref{eq:PPi.app} is identical to the fully Bayesian power formula in \eqref{eq:bayesianpo.app} with $n_i = n_o$, $n_j = n_r$ and $\hat\theta_i = \hat\theta_o$, the minimum PPi is simply $\Phi\left[\sqrt{t_i^2 - z_{\alpha/2}^2}\right]$.

\section{Package \texttt{R\lowercase{eplication}S\lowercase{uccess}}}\label{sec:package}
\subsection*{Installation}
\begin{knitrout}
\definecolor{shadecolor}{rgb}{0.969, 0.969, 0.969}\color{fgcolor}\begin{kframe}
\begin{alltt}
\hlcom{# install the package}
\hlkwd{install.packages}\hlstd{(}\hlstr{"ReplicationSuccess"}\hlstd{,}
                 \hlkwc{repos} \hlstd{=} \hlstr{"http://R-Forge.R-project.org"}\hlstd{)}
\hlcom{# load the package }
\hlkwd{library}\hlstd{(}\hlstr{"ReplicationSuccess"}\hlstd{)}
\end{alltt}
\end{kframe}
\end{knitrout}
\subsection*{Data}

\begin{knitrout}
\definecolor{shadecolor}{rgb}{0.969, 0.969, 0.969}\color{fgcolor}\begin{kframe}
\begin{alltt}
\hlcom{# load the data}
\hlkwd{data}\hlstd{(SSRP)}
\end{alltt}
\end{kframe}
\end{knitrout}

\begin{table}[!h]
\centering
\caption{\small Main variables of the \texttt{SSRP} dataset.}
\begin{tabular}{l l}
  \texttt{study} & Authors of the original paper \\
  \texttt{ro} & Original effect on correlation scale \\
  \texttt{ri} & Interim effect on correlation scale \\
  \texttt{rr} & Replication effect on correlation scale \\
  \texttt{fiso} & Original effect on Fisher-$z$ scale \\
  \texttt{fisi} & Interim effect on Fisher-$z$ scale \\
  \texttt{fisr} & Replication effect on Fisher-$z$ scale \\
  \texttt{se\_fiso} & Standard error of \texttt{fiso} \\
  \texttt{se\_fisi} & Standard error of \texttt{fisi} \\
  \texttt{se\_fisr} & Standard error of \texttt{fisr} \\
  \texttt{no} & Nominal sample size in original study \\
  \texttt{ni} & Nominal sample size in replication study at interim \\
  \texttt{nr} & Nominal sample size in replication study at the final analysis\\
  \texttt{po} & Two-sided $p$-value from original study \\
  \texttt{pi} & Two-sided $p$-value from interim analysis \\
  \texttt{pr} & Two-sided $p$-value from replication study 
\end{tabular}
\label{tbl:variables}
\end{table}
Table~\ref{tbl:variables} shows the main variables of the \texttt{SSRP} dataset.

\subsection*{Functions}
The two main functions that were used in this paper are \texttt{powerSignificance} and \texttt{powerSignificanceInterim}. The main arguments of these functions are presented in Table~\ref{tbl:arguments}.

\begin{table}[!h]
\centering
\caption{\small Arguments of the \texttt{powerSignificance} and \texttt{powerSignificanceInterim} functions. Arguments only used in \texttt{powerSignificanceInterim} are in grey.}
\begin{tabular}{l l}
\texttt{zo} & $z$-value from original study \\
\rowcolor{light-gray}
\texttt{zi} & $z$-values from interim analysis \\
\texttt{c} & 	Ratio of the replication sample size to the original sample size \\
\rowcolor{light-gray}
\texttt{f} & Fraction of the replication study already completed\\
\texttt{designPrior} & Design prior \\
\rowcolor{light-gray}
\texttt{analysisPrior} & Analysis prior \\
\texttt{alternative} & Direction of the alternative \\
\texttt{level} & Significance level \\
\texttt{shrinkage} & Shrinkage factor for original effect estimate
\end{tabular}
\label{tbl:arguments}
\end{table}

\newpage
\subsection*{Examples}
\begin{knitrout}
\definecolor{shadecolor}{rgb}{0.969, 0.969, 0.969}\color{fgcolor}\begin{kframe}
\begin{alltt}
\hlcom{# Variables}

\hlcom{# z-value from original study}
\hlstd{SSRP}\hlopt{$}\hlstd{zo} \hlkwb{=} \hlstd{SSRP}\hlopt{$}\hlstd{fiso}\hlopt{/}\hlstd{SSRP}\hlopt{$}\hlstd{se_fiso}

\hlcom{# z-value at interim}
\hlstd{SSRP}\hlopt{$}\hlstd{zi} \hlkwb{=} \hlstd{SSRP}\hlopt{$}\hlstd{fisi}\hlopt{/}\hlstd{SSRP}\hlopt{$}\hlstd{se_fisi}

\hlcom{# relative sample size from figure 4, }
\hlcom{# where stage 2 is ignored }
\hlcom{# (results at interim are considered }
\hlcom{# final results)}
\hlstd{SSRP}\hlopt{$}\hlstd{c_f3} \hlkwb{=} \hlstd{(SSRP}\hlopt{$}\hlstd{ni} \hlopt{-} \hlnum{3}\hlstd{)}\hlopt{/}\hlstd{(SSRP}\hlopt{$}\hlstd{no} \hlopt{-} \hlnum{3}\hlstd{)}

\hlcom{# same values can be found }
\hlcom{# with variance ratio}
\hlcom{# SSRP$c_f3 = SSRP$se_fiso^2/SSRP$se_fisi^2}

\hlcom{# relative sample size}
\hlstd{SSRP}\hlopt{$}\hlstd{c} \hlkwb{=} \hlstd{(SSRP}\hlopt{$}\hlstd{nr} \hlopt{-} \hlnum{3}\hlstd{)}\hlopt{/}\hlstd{(SSRP}\hlopt{$}\hlstd{no} \hlopt{-} \hlnum{3}\hlstd{)}

\hlcom{# same values with variance ratio}
\hlcom{# SSRP$c = SSRP$se_fiso^2/SSRP$se_fisr^2}

\hlcom{# fraction of replication study }
\hlcom{# already completed}
\hlstd{SSRP}\hlopt{$}\hlstd{f} \hlkwb{=} \hlstd{(SSRP}\hlopt{$}\hlstd{ni} \hlopt{-} \hlnum{3}\hlstd{)}\hlopt{/}\hlstd{(SSRP}\hlopt{$}\hlstd{nr} \hlopt{-} \hlnum{3}\hlstd{)}

\hlcom{# same result }
\hlcom{# with variance ratio}
\hlcom{# SSRP$f = SSRP$se_fisr^2/SSRP$se_fisi^2 }
\end{alltt}
\end{kframe}
\end{knitrout}

\begin{knitrout}
\definecolor{shadecolor}{rgb}{0.969, 0.969, 0.969}\color{fgcolor}\begin{kframe}
\begin{alltt}
\hlcom{# Example : calculation of the predictive power from Fig. 4}

\hlkwd{powerSignificance}\hlstd{(}\hlkwc{zo} \hlstd{= SSRP}\hlopt{$}\hlstd{zo,}
                  \hlkwc{c} \hlstd{= SSRP}\hlopt{$}\hlstd{c_f3,}
                  \hlkwc{designPrior} \hlstd{=} \hlstr{"predictive"}\hlstd{,}
                  \hlkwc{shrinkage} \hlstd{=} \hlnum{0.25}\hlstd{,}
                  \hlkwc{alternative} \hlstd{=} \hlstr{"two.sided"}\hlstd{,}
                  \hlkwc{level} \hlstd{=} \hlnum{0.05}\hlstd{)}
\end{alltt}
\end{kframe}
\end{knitrout}

\begin{knitrout}
\definecolor{shadecolor}{rgb}{0.969, 0.969, 0.969}\color{fgcolor}\begin{kframe}
\begin{alltt}
\hlcom{# Example: calculation of the informed predictive power }
\hlcom{# at interim (IPPi) from Table 3}

\hlcom{# select the studies that were continued }
\hlcom{# after the interim analysis}
\hlstd{SSRP_cont} \hlkwb{<-} \hlstd{SSRP[}\hlopt{!}\hlkwd{is.na}\hlstd{(SSRP}\hlopt{$}\hlstd{rr),]}

\hlkwd{powerSignificanceInterim}\hlstd{(}\hlkwc{zo} \hlstd{= SSRP_cont}\hlopt{$}\hlstd{zo,}
                         \hlkwc{zi} \hlstd{= SSRP_cont}\hlopt{$}\hlstd{zi,}
                         \hlkwc{c} \hlstd{= SSRP_cont}\hlopt{$}\hlstd{c,}
                         \hlkwc{f} \hlstd{= SSRP_cont}\hlopt{$}\hlstd{f,}
                         \hlkwc{designPrior} \hlstd{=} \hlstr{"informed predictive"}\hlstd{,}
                         \hlkwc{analysisPrior} \hlstd{=} \hlstr{"flat"}\hlstd{,}
                         \hlkwc{alternative} \hlstd{=} \hlstr{"two.sided"}\hlstd{,}
                         \hlkwc{level} \hlstd{=} \hlnum{0.05}\hlstd{)}
\end{alltt}
\end{kframe}
\end{knitrout}

\bibliographystyle{apalike}
\end{document}